\documentstyle[prb,aps,preprint]{revtex}
\begin{document}
\draft
\title{BCS and Generalized BCS Superconductivity in Relativistic Quantum Field Theory. II. Numerical Calculations}
\author{Tadafumi Ohsaku}
\address{Department of Physics, Graduate School of Science, Osaka University, Machikaneyama-cho 1-1, Toyonaka, Osaka, 560-0043, Japan}
\date{\today}
\maketitle

\newcommand{\bmx}{\mbox{\boldmath $x$}}
\newcommand{\bmy}{\mbox{\boldmath $y$}}
\newcommand{\bmk}{\mbox{\boldmath $k$}}
\newcommand{\bmp}{\mbox{\boldmath $p$}}
\newcommand{\bmq}{\mbox{\boldmath $q$}}
\newcommand{\bmP}{\mbox{\boldmath $P$}}  
\newcommand{\kfey}{\ooalign{\hfil/\hfil\crcr$k$}}
\newcommand{\pfey}{\ooalign{\hfil/\hfil\crcr$p$}}
\newcommand{\qfey}{\ooalign{\hfil/\hfil\crcr$q$}}
\newcommand{\Deltafey}{\ooalign{\hfil/\hfil\crcr$\Delta$}}  
\def\sech{\mathop{\rm sech}\nolimits}

\begin{abstract}

We solve numerically various types of the gap equations developed in the relativistic BCS and generalized BCS framework, presented in part I of this paper. We apply the method for not only the usual solid metal but also other physical systems by using homogeneous fermion gas approximation.  
We examine the relativistic effects on the thermal properties and the Meissner effect of the BCS and generalized BCS superconductivity of various cases.

\end{abstract}

\section{Introduction}

It is beyond doubt that the relativistic effects are important in condensed matter physics: They are present in various systems. But until recently, not so much investigation for relativistic effects in the phenomena of superconductivity were performed, except for taking into account the spin-orbit coupling~[1]. Recently, the papers of Capelle et al. appeared, in which the necessity of the relativistic treatment for superconductivity in condensed matter was asserted~[2$\sim$4]. In the study of the superconductivity under strong relativistic effects, they argued the necessity of the full relativistic treatment, beyond that of making small corrections like the spin-orbit coupling. They introduced the Dirac-type Bogoliubov-de Gennes theory and discussed some new aspects of the relativistic superconductivity. Based upon their works, we try to generalize their methods in this paper.  

In part I of this paper, we developed the formalism of the BCS and generalized BCS superconductivity in relativistic quantum field theory~[5]. By using the Gor'kov formalism, we derived various types of gap equations. 
We also gave the group-theoretical consideration of the pairing gap functions. 
We treated not only the spin singlet pairing states, but also the spin triplet finite angular momentum pairing states by introducing the relativistic generalized-BCS scheme. 

We try now to complete our program and work out the numerical study of the gap equations for various cases, by using the formalism developed in part I. To examine the relativistic effects on the gap function, thermodynamics and electromagnetic response, we have to study various cases by performing numerical calculations. In Sec. II, we examine the excitation energy spectra of quasiparticles in the relativistic theory. This examination will be useful for investigation and interpretation of the relativistic gap equations and the thermodynamics of the superconductor. In Sec. III, we solve various types of gap equations, given in part I of this paper. We consider not only the situation of the usual solid metal, but also other situations where the system has a larger Fermi energy. {\it By studying these systems, we depict the strength of the relativistic effects on the gap equations under an unified perspective.} Because we choose the theoretical framework as the BCS and generalized BCS theory, we can materialize it. In Sec. IV, the thermodynamics and the Meissner effect in the relativistic BCS and generalized BCS superconductivity are discussed. Finally, in Sec. V, we give a summary of the present investigation and provide further possible extensions to be studied for the pairing properties of matter.

\section{The Excitation energy spectra of Quasiparticles and The Density of States}

In this section, we examine the excitation energy spectra of the quasiparticles, and the density of states (DOS). Especially we concentrate on the scalar and vector pairings, $\Delta^{S}$ and $\Delta^{V}_{0}$~[5]. As discussed in part I~[5], the pairing gap function $\Delta_{4\times 4}=\langle\psi\psi^{T}\rangle$ is decomposed by 16-dimensional complete set of $\gamma$ matrices. After constructing the gap equation for each component of them, we found the fact that, only $\Delta^{S}$, $\Delta^{V}_{0}$, ${\bf \Delta}^{A}$ ( the spacelike components of axial vector pairing ) and ${\bf \Delta}^{T}_{(A)}$ ( the axial-vector-like components of the two-rank antisymmetric tensor pairing ) can have nontrivial solutions. Under the treatment given in part I, in all the cases of ${\bf \Delta}^{A}$ and ${\bf \Delta}^{T}_{(A)}$, the energy spectra of the quasiparticles of these states take the similar forms with that of $\Delta^{S}$ or $\Delta^{V}_{0}$, except the differences about the angular dependences of the gap functions in momentum space. Thus, we regard the energy spectra of $\Delta^{S}$ and $\Delta^{V}_{0}$ as having the typical character of the quasiparticles in the relativistic BCS and generalized BCS theory. 

First, we give the excitation energy spectra of the quasiparticles of the $\Delta^{S}$ and $\Delta^{V}_{0}$ states schematically in Fig. 1. $E_{+} ( E_{-} )$ is the branch of the quasiparticles coming from positive ( negative ) energy states. The explicit forms are given as
\begin{eqnarray}
E_{\pm} &=& \sqrt{(\sqrt{\bmk^2+m^2}\mp\mu)^{2}+|\Delta^{S}|^2},
\end{eqnarray}
for the scalar pairing~[2,5], and
\begin{eqnarray}
E_{\pm} &=& \sqrt{\bmk^{2}+m^{2}+\mu^{2}+|\Delta^{V}_{0}|^{2}\mp 2\sqrt{(\bmk^{2}+m^{2})\mu^{2}+|\Delta^{V}_{0}|^{2}\bmk^{2}}},
\end{eqnarray}
for the 0th component of the vector pairing~[5]. We consider first the spectrum $E_{+}$. In the case of $\Delta^{S}$, the minimum is found at $k_{min}=k_{F}=\sqrt{\mu^{2}-m^{2}}$ ( Throughout this paper, we take the approximation to set $\mu=\epsilon_{F}$. This is the case for zero temperature. We completely neglect the temperature dependence of $\mu$ and use it as a simple parameter to give the particle density of a system. ), and the energy gap is given by the width $2|\Delta^{S}|$, same as the usual nonrelativistic case. On the other hand, in the case of $\Delta^{V}_{0}$, the minimum is found at $k_{min}=\sqrt{\mu^{2}+|\Delta^{V}_{0}|^{2}-m^{2}\mu^{2}/(\mu^{2}+|\Delta^{V}_{0}|^{2})} ( > k_{F} )$, and the gap width becomes $2m|\Delta^{V}_{0}|/\sqrt{\mu^{2}+|\Delta^{V}_{0}|^{2}} ( < 2|\Delta^{V}_{0}| )$. The gap exists also at $k_{F}$, but particles feel smaller gap at $k_{min}$. Usually $\mu\gg|\Delta^{V}_{0}|$, and therefore the gap width becomes $\sim 2\frac{m}{\mu}|\Delta^{V}_{0}|$. For $\frac{\mu-m}{m}\ll 1$ ( like the case of usual solid metal ), the gap width becomes nearly $2|\Delta^{V}_{0}|$. In this case, the spectra of $\Delta^{S}$ and $\Delta^{V}_{0}$ almost coincide with each other. It is also the case that, when $\mu-m$ becomes large, the difference of the spectra between $\Delta^{S}$ and $\Delta^{V}_{0}$ becomes large. Futher, it is clear from the above discussion, if we treat the massless case, the gap will vanish in the $\Delta^{V}_{0}$ state. 

The branch of $E_{-}$ is located above $E_{+}$ in both cases. When $\frac{\mu-m}{m}\ll 1$, the relative difference between $E_{+}$ and $E_{-}$ is large compared with the Fermi energy, $\mu -m$ ( which gives the characteristic energy scale of a system ). Hence, the contribution of the quasiparticles of $E_{-}$ becomes small or can be neglected. On the other hand, when $\mu-m \sim m$ ( like the relativistic plasma ), the contribution of $E_{-}$ would become large, and we should take $E_{-}$ into consideration for the pairing properties. 

The density of states ( DOS ) is estimated by 
\begin{eqnarray}
\rho_{s}(\omega) &=& \rho_{n}(\epsilon_{F})\int d\xi \delta(\omega-E_{+}) \nonumber \\
&=& \rho_{n}(\epsilon_{F})\int dE_{+}\frac{d\xi}{dE_{+}}\delta(\omega-E_{+}),
\end{eqnarray}
where $\rho_{s}(\omega)$ is the DOS of the superconducting state, while $\rho_{n}(\epsilon_{F})$ is the DOS of the normal state at the Fermi energy; $\rho_{n}(\epsilon_{F})=\mu\sqrt{\mu^{2}-m^{2}}$, and $\xi\equiv\sqrt{k^{2}+m^{2}}-\mu$. Here we only take into account the contribution of the branch $E_{+}$. Fig. 2 schematically depicts the DOS for the $\Delta^{S}$ and $\Delta^{V}_{0}$ states. In the case of $\Delta^{S}$, $\rho_{s}(\omega)$ approaches to infinity at $\omega\to\pm|\Delta^{S}|$, while in $\Delta^{V}_{0}$ state, $\rho_{s}(\omega)$ becomes infinity at $\omega\to\pm m|\Delta^{V}_{0}|/\sqrt{\mu^{2}+|\Delta^{V}_{0}|^{2}}$. Therefore, when $|\Delta^{S}|=|\Delta^{V}_{0}|$, the quasiparticles in the $\Delta^{V}_{0}$ state is more easily excited than that of the $\Delta^{S}$ state.

\section{Numerical Solutions of The Gap Equations}

In this section, we discuss the numerical solutions of the gap equations. In the numerical calculations, we should not use the approximation which is usually used in the condensed matter theory:
\begin{eqnarray}
\int\frac{d^{3}\bmk}{(2\pi)^{3}} &\to& \rho_{n}(\epsilon_{F})\int d\xi. 
\end{eqnarray}
Rather, we have to treat the details of the energy spectra in the integration of the gap equation in order to compare the relativistic and nonrelativistic cases. Furthermore, we have to introduce a cutoff to make the integral of the gap equation finite due to the use of the $\delta$-function expression for the interaction for simplicity~[5]. We take the energy cutoff $\Lambda$ in the following way:
\begin{eqnarray}
\int\frac{d^{3}\bmk}{(2\pi)^{3}} \to \frac{1}{(2\pi)^{3}}\int^{\epsilon_{F}+\Lambda}_{\epsilon_{F}-\Lambda}\epsilon\sqrt{\epsilon^{2}-m^{2}}d\epsilon\int^{\pi}_{0}\sin\theta d\theta\int^{2\pi}_{0} d\phi.
\end{eqnarray}
As discussed in part I, the aim of this work is to study the relativistic BCS and generalized BCS theory. For this aim, we set aside the question of the mechanism of the origin of the superconductivity. This is the attitude of the BCS and generalized BCS theory. Due to this attitude, we can study the relativistic superconductivity of various systems in an unified approach, from the usual solid to the stellar matter. Therefore, we treat the cutoff in a general way. In the electron-phonon interaction, we only treat a thin shell near the Fermi surface. Since we skip the question about the origin of the attractive interaction, we can extend the width of integration as $T_{Q}\le\Lambda\le T_{F}$, based on the discussion of the textbook of Vollhardt and W\"{o}lfle~[6], chapter 3. Here $k_{B}T_{Q}\equiv 1/\tau$ ( $\tau$; the quasiparticle lifetime of the Fermi liquid ), while $T_{F}$ means the Fermi temperature. We integrate the states of the inside and outside of the Fermi sphere symmetrically, with moving the cutoff $\Lambda$ in this range. Then, the gap equation generally has the cutoff $\Lambda$, coupling constant $g_{l}$, fermion mass $m$, chemical potential $\mu$ and temperature $T$ as the parameters of the model.  

Our gap equations have five parameters. Therefore, there is a large arbitrariness to choose these values. In this work, we mainly investigate the relativistic effects in superconductivities of several physical systems. To determine the conditions of solving the gap equations, we present the typical order of physical quantities of several systems, in table I. In this table, we consider 5 examples: (i) The usual solid metal ( metallic elements, alloys, compounds, oxides and organic metals )~[7,8], (ii) the metallic hydrogen~[8,9,10], (iii) to depict the strength of relativistic effects in heavy elements, we introduce an imaginary system of electron gas, in which $\epsilon_{F}-m$ is almost same as the kinetic energy of $1s$-electron of uranium atom, (iv) the relativistic plasma~[11] and (v) the neutron star~[12]. To estimate the Fermi energy $\epsilon_{F}$ of the systems, we use the nonrelativistic energy-momentum relation $\epsilon_{F}=m+\frac{k^{2}_{F}}{2m}$ for the cases of (i), (ii) and (v), while we use the relativistic energy-momentum relation $\epsilon_{F}=\sqrt{k^{2}_{F}+m^{2}}$ for the cases of (iii) and (iv). For (ii), the value of $r_{s}\equiv\frac{r_{0}}{a_{B}}$ ( $r_{0}$; mean particle distance, $a_{B}$; the Bohr radius ) takes the range of $0.5\le r_{s}\le 1.5$~[9,10]. Based on the paper of Ichimaru~[10], the electron system is relativistic at $r_{s} < 0.1$ with $k_{B}T/(\epsilon_{F}-m) \ll 0.1$ ( at $r_{s}=0.1$, $\epsilon_{F}-m = 5\times10^{3}$eV and $\frac{\epsilon_{F}-m}{m}\sim10^{-2}$ ). For (iii), we take the mean radius of $1s$-electron as $\langle r\rangle_{1s}=715\times10^{-13}$cm~[13], and estimate the particle density by $(\frac{4}{3}\pi\langle r\rangle^{3}_{1s})^{-1}$. In this case, we obtain $r_{s}=0.013$. It is known that the relativistic effect ( kinetic energy effect ) enhances the screening, and the screening length takes a finite value at $r_{s}\to 0$, while the nonrelativistic Thomas-Fermi length diverges in the same limit~[10,14]. Therefore, we shall consider the superconductivity in such high density matter, with an assumption of existence of some effective attractive interaction. 

Consulting with table I, we put the values of $\epsilon_{F}$ to $\mu$ in the gap equations. Commonly, the magnitude of the gap is much smaller than $\mu -m$. For example, in usual solid metal, the ratio $|\Delta(T=0)|/(\mu-m)$ becomes $10^{-3}\sim 10^{-4}$. Hence, we search the values of our parameters, especially $g_{l}$ and $\Lambda$, to satisfy this condition in solving the gap equations. After choosing the values of the parameters, the integration in the gap equation is performed, and search the self-consistency condition under the variation with respect to the amplitude of the gap. For the integration, we used the numerical package {\sl Mathematica} version 4.1. Usually, this package gives high numerical accuracy. 

First, we treat the spin singlet pairing states. We take the cases of $\Delta^{S}$ ( scalar ):
\begin{eqnarray}
1 &=& \frac{g_{0}}{4\pi^{2}}\int^{\mu+\Lambda}_{\mu-\Lambda}d\epsilon\epsilon\sqrt{\epsilon^{2}-m^{2}}\Bigl(\frac{1}{2E_{+}}\tanh\frac{\beta}{2}E_{+} + \frac{1}{2E_{-}}\tanh\frac{\beta}{2}E_{-}\Bigr), \\
E_{\pm} &=& \sqrt{(\epsilon\mp\mu)^{2}+|\Delta^{S}|^{2}},
\end{eqnarray}
$\Delta^{V}_{0}$ ( 0th component of vector ):
\begin{eqnarray}
1 &=& \frac{g_{0}}{4\pi^{2}}\int^{\mu+\Lambda}_{\mu-\Lambda}d\epsilon\epsilon\sqrt{\epsilon^{2}-m^{2}}\Bigl(\{1-\frac{\epsilon^{2}-m^{2}}{\sqrt{\epsilon^{2}\mu^{2}+|\Delta^{V}_{0}|^{2}(\epsilon^{2}-m^{2})}}\}\frac{1}{2E_{+}}\tanh\frac{\beta}{2}E_{+} \nonumber \\
 & & +\{1+\frac{\epsilon^{2}-m^{2}}{\sqrt{\epsilon^{2}\mu^{2}+|\Delta^{V}_{0}|^{2}(\epsilon^{2}-m^{2})}}\}\frac{1}{2E_{-}}\tanh\frac{\beta}{2}E_{-}\Bigr),  \\
E_{\pm} &=& \sqrt{\epsilon^{2}+\mu^{2}+|\Delta^{V}_{0}|^{2}\mp2\sqrt{\epsilon^{2}\mu^{2}+|\Delta^{V}_{0}|^{2}(\epsilon^{2}-m^{2})}}, 
\end{eqnarray} 
``no-sea'' ( neglect the contribution from the negative energy states in $\Delta^{S}$ ):
\begin{eqnarray}
1 &=& \frac{g_{0}}{4\pi^{2}}\int^{\mu+\Lambda}_{\mu-\Lambda}d\epsilon\epsilon\sqrt{\epsilon^{2}-m^{2}}\frac{1}{2E_{+}}\tanh\frac{\beta}{2}E_{+}, \\
E_{+} &=& \sqrt{(\epsilon-\mu)^{2}+|\Delta^{S}|^{2}},
\end{eqnarray}
and ``nonrelativistic'' ( substitute $\sqrt{\bmp^{2}+m^{2}}\to m+\frac{\bmp^{2}}{2m}$ to ``no-sea'' ):
\begin{eqnarray}
1 &=& \frac{g_{0}}{4\pi^{2}}\int^{\mu+\Lambda}_{\mu-\Lambda}d\epsilon\sqrt{2m^{3}(\epsilon-m)}\frac{1}{2E}\tanh\frac{\beta}{2}E, \\
E &=& \sqrt{(\epsilon-\mu)^{2}+|\Delta^{NR}|^{2}}.
\end{eqnarray}
Therefore, we treat these 4-cases of the gap equations for comparison.

Fig. 3 gives the gap functions at $T=0$ as a function of the coupling constant $g_{0}$. Here we take the cutoff $\Lambda$ such that the integration is performed over the inside and the outside of the Fermi sphere symmetrically; $\Lambda/(\mu-m)=1$.  In the case of the solid, the solutions obtained from these 4-types of gap equations discussed above coincide with each other very well. In this case, the relativistic effect is negligible at least in the gap function. There is no contribution of the quasiparticles coming from the negative energy states. In fact, the gap equations of $\Delta^{S}$ and $\Delta^{V}_{0}$, Eqs. (6) and (8) almost coincide in the limit $\frac{\mu-m}{m}\ll 1$ with $|\Delta|\ll\mu$. On the other hand, in the case of the relativistic plasma, there are large differences between these 4 solutions. $\Delta^{S}$ gives the largest gap, while $\Delta^{V}_{0}$ is strongly suppressed and becomes the smallest gap. The gap of the ``nonrelativistic'' case is also largely reduced compared with $\Delta^{S}$. The difference of $\Delta^{S}$ and ``no-sea'' is also clear. At $g_{0}=15.8$MeV$^{-2}$, the relative ratios are 105 for $\Delta^{S}$/``nonrelativsitic'' and 81.4 for ``no-sea''/``nonrelativistic''. Therefore, in the relativistic plasma, we have to introduce the relativistic treatment, and cannot neglect the contribution of the negative energy states. For the case of the neutron star, the solutions of the 4-cases are close from each other. But we observe small differences between them. At $g_{0}=3.55\times10^{-5}$MeV$^{-2}$, the relative ratios are 1.16 for $\Delta^{S}$/``nonrelativistic'', and 1.15 for ``no-sea''/``nonrelativistic''. This indicates that the contribution of the negative energy states to the gap function is less than 1 percent in this case. It is clear from Fig. 3, the solutions reflect the nonperturbative effect and behave almost $e^{-1/g_{0}}$ in all cases. 

In Fig. 4, we show the dependence of the cutoff $\Lambda$, namely the integration width, in the solutions of 4-cases at $T=0$. Here, $x\equiv 2\Lambda$. We squeeze the width to a small region containing Fermi surface. At $\Lambda/(\mu-m)=1$, we integrate all the inner region of the Fermi sphere and outer region symmetrically. Same as in Fig. 3, the solutions of the 4-cases coincide very well in the case of the solid, and $|\Delta|\ll\Lambda$ is always retained. In the cases of the relativistic plasma, when we squeeze the cutoff, the gaps of $\Delta^{S}$ and ``no-sea'' come close to each other. But the relative ratio between $\Delta^{S}$ and ``nonrelativistic'' still remains large. We also observe that in the case of the neutron star, the solutions of the 4-cases come close to each other when we squeeze the cutoff. In all cases, the solutions almost linearly depend on $x$. Therefore, we conclude that the solutions behave almost $|\Delta(T=0)|\propto \Lambda e^{-1/g_{0}}$ ( like the nonrelativistic BCS theory~[8,15,16] ) in 4-cases of various systems. 

In Figs. 3 and 4, the values of $|\Delta(T=0)|$ are given almost in the same range. From the results of Figs. 3 and 4, we find the values of $g_{0}$ and $\Lambda$ to provide reasonable magnitude of the gap, and then we solve the $T$-dependent gap equations. Fig. 5 gives the temperature $T$ dependences of the solutions in the 4-cases. For the solid, the solutions of the 4-cases almost coincide. The critical temperature $T_{c}$ of Fig. 5(a) is 20.7 K. The case of metallic hydrogen ( here we take $r_{s}=0.5$ ), we find a tiny difference between $\Delta^{S}$ and ``nonrelativistic'' cases, and $T_{c}\sim$750 K. In the case of the relativistic plasma, $\Delta^{V}_{0}$ and ``nonrelativistic'' are largely reduced. The difference of $\Delta^{S}$ and ``no-sea'' is also large. $T_{c}$ of $\Delta^{S}$ is 0.0820 MeV = 9.5$\times$10$^{8}$ K. This $T_{c}$ is 0.16 times the electron rest mass. For the neutron star, $T_{c}$ of Fig. 5 for $\Delta^{S}$ is 1.86 MeV, while 1.71 MeV for the nonrelativistic case; thus the magnitude of the relativistic effect is less than 10 percent. 

Next, to describe our numerical results of the singlet states of various electron systems in an unified manner, we solve the gap equations with very fine variation of the chemical potential. We show the results in Fig. 6. The value of $g_{0}$ is determined so as to provide the ratio $|\Delta(T=0)|/(\mu-m)=10^{-3}$ in the solution of the ``nonrelativistic'' case obtained from Eq. (12). The cutoff is always set as $\Lambda/(\mu-m)=1$. By using these values for $g_{0}$ , we solve the gap equations (6), (8), (10) and (12), and obtain the values of $|\Delta(T=0)|$ and $T_{c}$. The deviations from ``nonrelativistic'' give the relativistic effects. The relativistic effects become significant for $\mu-m\sim 10$keV in both quantities. At $\mu-m=0.5$MeV, $T_{c}$ for $\Delta^{S}$ case becomes 0.077 times the electron rest mass. We obtain an important result that, by using the data in Fig. 6, the BCS universal constant $|\Delta(T=0)|/k_{B}T_{c}$=1.76~[8,15,16] is almost completely filled in $\Delta^{S}$, ``no-sea'' and ``nonrelativistic'' in all $\mu-m$, while discrepancies are obtained at $\mu-m > 50$keV for $\Delta^{V}_{0}$ state. Thus, we conclude that $\Delta^{S}$, ``no-sea'' and ``nonrelativistic'' cases obey the BCS-like temperature dependence in all situations.  

From the above results, we conclude that if the chemical potential $\mu$ is large enough, generally the gaps of the $\Delta^{S}$ case and ``no-sea'' are larger than that of the nonrelativistic case, and $T_{c}$ also becomes higher. Especially in the case when the attractive interaction requires the treatment covering the wide range of inner and outer regions of the Fermi sphere, those relative differences become significant. It is well known that, the DOS of the relativistic theory in normal state $\epsilon\sqrt{\epsilon^{2}-m^{2}}$ is larger than that of the nonrelativistic one $\sqrt{2m^{3}(\epsilon-m)}$. This fact also contributes to enhance the solutions of the relativistic gap equations as compared with nonrelativistic counterpart. The relative difference between the ``no-sea'' and ``nonrelativistic'' cases has its origin in this effect. We expect from the relative relation among the magnitudes of the gap, the condensation energy, the lowering of the free energy and the thermodynamic critical magnetic field become larger in the relativistic case than in the nonrelativistic case. When $(\mu-m)/m\ll 1$ ( like the usual solid metal ), we cannot distinguish $\Delta^{S}$ and $\Delta^{V}_{0}$. In such a case, the presence of the $\Delta^{V}_{0}$ state is ``hidden''. But actually, $\Delta^{S}$ and $\Delta^{V}_{0}$ can be mixed under $O(3)$ rotational symmetry ( as discussed in part I of this paper ).   

The reason of the relation $|\Delta^{S}|\ge|\Delta^{V}_{0}|$ in all $\mu$ is understood in the following way. Based on Sec. III of part I, each mean field is given as follows:
\begin{eqnarray}
\Delta^{S} &\propto& \langle \psi^{T}(-C\gamma_{5})\psi\rangle, \\
\Delta^{V}_{0} &\propto& \langle \psi^{T}(-C\gamma_{5})\gamma_{0}\psi\rangle. 
\end{eqnarray}
Here, $T$ means the transposition of a matrix. In the standard representation, $\psi$ is given as
\begin{equation}
\psi = \left(
\begin{array}{c}
\varphi \\
\chi
\end{array}
\right),
\end{equation}
where $\varphi$ is the large component, while $\chi$ is the small component. Thus, from the relations of (14) and (15), we obtain
\begin{eqnarray}
\Delta^{S} &=& \varphi^{T}(i\sigma_{2})\varphi + \chi^{T}(i\sigma_{2})\chi, \\
\Delta^{V}_{0} &=& \varphi^{T}(i\sigma_{2})\varphi - \chi^{T}(i\sigma_{2})\chi.
\end{eqnarray} 
Here $i\sigma_{2}$ expresses the spin singlet pairings clearly. At $\mu\to m$, the small component vanishes, and $\Delta^{S}$ and $\Delta^{V}_{0}$ coincide. On the other hand, when $\mu-m\sim m$ and the kinetic energy becomes large, $\varphi$ and $\chi$ approach to the same limit ( when $m/\mu \to 0$, they coinside with each other ), then the difference between the magnitudes of the $\Delta^{S}$ and $\Delta^{V}_{0}$ becomes large. ( We can choose the phase $\theta=0$ in $\Delta^{S}=|\Delta^{S}|e^{i\theta}$ and $\Delta^{V}_{0}=|\Delta^{V}_{0}|e^{i\theta}$. )

In this work, we do not present the results of the massless case and/or ultrarelativistic limit. We mention, however, for the case of $\Delta^{V}_{0}$, we obtain an unphysical result at $m=0$; two solutions appear in our numerical calculation. 

Next we give the results for the spin triplet gap equations in Fig. 7. We solve our gap equations, given in (114)$\sim$(117) with table I of part I, under the 3 conditions: (a) The usual solid metal, (b) the relativistic plasma and (c) the neutron star. In all these cases, at $T=0$, the largest solutions are obtained for the $\psi_{000}$ state of ${\bf \Delta}^{T}_{(A)}$. This state is regarded as the Balian-Werthamer (BW) state~[6,8,17] in our theory, as discussed in part I of this paper~[5]. Thus our theory also predicts that the BW state is the most important state, like the nonrelativistic theory. The $\psi^{(-)}_{101}$ state of ${\bf \Delta}^{A}$ ( the Anderson-Brinkman-Morel (ABM) state~[6,8,18] in our theory~[5] ) gives smaller solutions than that of the BW state at $T=0$. The $T_{c}$ of the BW, ABM and $\frac{1}{\sqrt{2}}(\psi^{(-)}_{111}-\psi^{(-)}_{1-11})$ of ${\bf \Delta}^{A}$ coincide very well. In all 3 conditions given above ( (a), (b) and (c) ), the values of $|\Delta(T=0)|/k_{B}T_{c}$ are 6.25 for the BW, while 5.87 for the ABM. Especially for the case of ${\bf \Delta}^{T}_{(A)}\psi_{000}$ and ${\bf \Delta}^{A}\psi_{000}$, their relations to each other in the gap equations are the same as that for $\Delta^{S}$ and $\Delta^{V}_{0}$ except for the coefficients. Therefore, to interprete the results of the spin triplet gap equations, we can use the main results of the spin singlet states. We can conclude that, when $(\mu-m)/m\ll 1$, the solutions of ${\bf \Delta}^{A}$ and ${\bf \Delta}^{T}_{(A)}$ in the same $\psi_{jm_{j}\lambda}$ will coincide very well. On the other hand, when $\mu-m\sim m$, the cases of the transversal states of ${\bf \Delta}^{A}$ and the longitudinal states of ${\bf \Delta}^{T}_{(A)}$ remain important states; the solutions of other cases given in part I will be strongly suppressed and become unimportant. Thus we recognize the fact that, in the relativistic theory, the relative orientation between the spin vector and momentum vector ( it is given by the helicity of the Cooper pair, as discussed in part I ) determine the magnitude of the stabilization of the superconductivity. Under the variation of parameters ( $g_{1}$, $\Lambda$ and $\mu$ )of the gap equations, the transversal states of ${\bf \Delta}^{A}$ and the longitudinal states of ${\bf \Delta}^{T}_{(A)}$ behave like $\Delta^{S}$, while other cases given in part I behave like $\Delta^{V}_{0}$. It is also the case that, in the cases of $j=2$ pairings, the solution of their states becomes relatively much smaller than those of the BW and ABM states. These results come from the fact that the basis $\psi_{jm_{j}\lambda}(j=2)$ for expanding these states has components which do not couple with p-wave interaction $g_{1}\sum Y_{1m}Y^{*}_{1m}$, because of the orthogonality coming from the azimuthal angular dependence. If we extend our theory to include f-wave components in the pairing interaction, these situations may become different. 

To summarize the numerical results of this section, we would like to mention that it is impossible to remove the arbitrariness of the choice of the values of the model parameters completely. Especially, the magnitudes of the solutions of the gap equations are sensitive to the coupling constants and the cutoff parameters. Therefore in our work, we can only discuss the characteristic features of the physics of the relativistic BCS superconductivity. We assert, however, that the various qualitative features we have obtained are true and we have to keep in mind the importance of the relativistic effects in some situations discussed above. In such situations, the results obtained here should be considered seriously.

\section{The Thermodynamics and The Meissner effect}

In this section, we discuss the thermodynamics, especially the specific heat and spin paramagnetic susceptibility, and the Meissner effect. First, we consider the specific heat of the superconducting state. It is well-known in the nonrelativistic theory~[6,8,19] that the temperature dependence of the specific heat $C$ at $T\ll T_{c}$ is determined by the DOS at the Fermi surface. The DOS of the Fermi surface is reflected by the node structure of the gap function. If the gap has no node, the specific heat depends exponentially on $T$ ( like the usual BCS or the BW state ), while $C\propto T^{2}$ for the case when the gap vanishes at a line ( the polar state ) and $C\propto T^{3}$ for the case when the gap vanishes at two points ( the ABM state ). To estimate the $T$-dependence of the specific heat in the relativistic theory, we use the next formula:
\begin{eqnarray}
C \propto \int\frac{d^{3}\bmk}{(2\pi)^{3}}\frac{1}{T^{2}}\Bigl(E^{2}_{+}\sech^{2}\frac{\beta}{2}E_{+}+E^{2}_{-}\sech^{2}\frac{\beta}{2}E_{-}\Bigr).
\end{eqnarray}
( for all the details of the derivation, see appendix A )
Thus, $C$ is determined by DOS, $T$, $E_{+}$, $E_{-}$ and the derivatives of the Fermi distribution functions. The second term is usually negligible. When $(\mu-m)/m \ll 1$, the excitation spectra of $\Delta^{S}$ and $\Delta^{V}_{0}$ almost coincide. On the other hand, when $\mu$ becomes large, the difference between the excitation spectra of $\Delta^{S}$ and $\Delta^{V}_{0}$ becomes significant, as discussed in Sec. II. In such a case, the quasipaticles are more easily excited in the $\Delta^{V}_{0}$ pairing than in the $\Delta^{S}$ pairing. But in this situation, $\Delta^{V}_{0}$ becomes unimportant, because in such a case, the solution of the gap equation of the $\Delta^{V}_{0}$ becomes small compared with $\Delta^{S}$, ``no sea'', ``nonrelativistic'', as discussed in Sec. III. Therefore, the effect of the difference in the excitation spectra of the radial direction in $\bmk$-space ( like the difference of $E_{+}$ in $\Delta^{S}$ and $\Delta^{V}_{0}$ ) does not affect the thermodynamics seriously. We conclude that, in the relativistic theory, we can also estimate the $T$-dependence of the specific heat by the node structure of the gap. Therefore, $\Delta^{S}$, $\Delta^{V}_{0}$, BW and $\frac{1}{\sqrt{2}}(\psi^{(-)}_{111}-\psi^{(-)}_{1-11})$ of ${\bf \Delta}^{A}$ of our theory have the exponential dependence, while the ABM state of our theory has $T^{3}$-dependence. In our theory, there is no state which has the polar state like node structure ( see, part I of this paper ). 

Next, we consider the spin paramagnetic susceptibility. Here we only consider $\Delta^{S}$, $\Delta^{V}_{0}$ and ``nonrelativistic'' cases; the spin singlet states. We introduce the Zeeman energy for a weak homogeneous external magnetic field, and employ the usual treatment, obtaining the magnetization 
\begin{eqnarray}
M &=& -\mu_{B}\int\frac{d^{3}\bmk}{(2\pi)^3}\biggl\{[f(E_{+}+\mu_{B}H)-f(E_{+}-\mu_{B}H)]+[f(E_{-}+\mu_{B}H)-f(E_{-}-\mu_{B}H)]\biggr\}.  \nonumber \\
 & &
\end{eqnarray}
From this, the spin paramagnetic susceptibility becomes
\begin{eqnarray}
\chi_{s} &=& -2\mu^{2}_{B}\int\frac{d^{3}\bmk}{(2\pi)^{3}}(\frac{\partial f(E_{+})}{\partial E_{+}}+\frac{\partial f(E_{-})}{\partial E_{-}}).
\end{eqnarray}
Here the second term is negligible. When we neglect this term, the difference of $\Delta^{S}$ and ``nonrelativistic'' is only the DOS. From this observation, we suppose there is no qualitative difference in the $T$-dependences in $\Delta^{S}$ and ``nonrelativistic'' cases. When $\mu-m\sim m$, the DOS enhances $\chi_{s}$ of $\Delta^{S}$, but the gap also becomes large compared with that of the ``nonrelativistic'' pairing, and thermal excitation becomes more difficult in $\Delta^{S}$ than in ``nonrelativistic''. In this situation, $|\Delta^{V}_{0}|$ is largely reduced and the $\Delta^{V}_{0}$ state becomes unimportant ( as discussed in Sec. III ). To retain the comparison of $\Delta^{S}$, $\Delta^{V}_{0}$ and ``nonrelativistic'' meaningful, we have to take $\mu-m$ small enough or a moderate value. In such situations ( the usual solid metal, metallic hydrogen and neutron star ), the DOS and $E_{+}$ of them are almost same values. In fact, we obtain numerical results that, in such situations, $\Delta^{S}$, $\Delta^{V}_{0}$ and ``nonrelativistic'' cases give the same $T$-dependences for $\chi_{s}$, and behave like the usual Yosida function~[8,20]. We also confirm in our numerical calculation, at large $\mu$ ( relativistic plasma ), the $\chi_{s}$ of $\Delta^{S}$ also behaves like the Yosida function ( at large $\mu-m$, it is needless to consider the $\Delta^{V}_{0}$ state ), and the second term in (21) is completely negligible. 

Strictly speaking, the treatment given above is not completely a relativistic one, because the Zeeman energy is a first-order term in $1/c$ ( here $c$ is the velocity of light ). Here, we only want to see the effect of the magnetic field in the simple method, which is usually used in the nonrelativistic theory. To compensate this incompleteness, we calculate the electromagnetic response function, and examine the Meissner effect, especially for the $\Delta^{S}$ and $\Delta^{V}_{0}$ states. The derivation of the response function is given in appendix B. We find, at a small or a moderate value of $\mu-m$ ( the usual solid metal, metallic hydrogen and neutron star ), the $T$-dependence of the numerical integrals of (B17) become almost $1-(T/T_{c})^{4}$ in the $\Delta^{S}$ and $\Delta^{V}_{0}$ states, similar to the behavior of the response in the nonrelativistic theory~[16]. This behavior is found also for the case of large $\mu-m$ ( the relativistic plasma ) for $\Delta^{S}$ pairing.

In summary, we conclude that, in the $T$-dependence of several thermodynamic quantities and Meissner effect, there is no qualitative difference between the relativistic BCS theory and nonrelativistic BCS theory in homogeneous systems. Therefore, we can safely use the basic concepts they are used in conventional nonrelativistic BCS and generalized BCS theory of superconductivity, for qualitative understanding of several thermodynamics and Meissner effect in various systems, from the usual solid metal to the stellar matter. This fact is not clear before our work is done.

\section{Concluding Remarks}

In part I and part II of this paper, we have performed the investigation of the BCS and generalized BCS superconductivity in relativistic quantum field theory. After the preparation of the Gor'kov equation and the group theoretical consideration of the pairing gap functions, we have solved the Gor'kov equations under the conditions which are assumed to have specific types of the pairing gap functions. After that, we have constructed various types of gap equations, and have solved them numerically. We have discussed the results in detail. The thermodynamic functions, especially the specific heat and the spin susceptibility were discussed. Finally we have examined the Meissner effect in our theory. We have obtained and examined various characteristic features of the relativistic BCS and generalized superconductivity throughout this paper.

In part II, we have mainly investigated the kinetic energy effect of the homogenous systems in the relativistic superconductivity. This effect clearly appears in the gap equations of several situations. On the other hand, the works of Capelle et al.~[2$\sim$4] treated the systems under deep periodic potentials. They considered the inhomogeneous effect in the relativistic superconductivity, and new physics they have obtained are based essentially on this effect. 

Now we discuss further possible extensions. In the case of the vector ${\bf \Delta}^{A}$ or tensor ${\bf \Delta}^{T}_{(A)}$ pairings, we have treated only the unitary cases. These treatments should be extended to study the case of nonunitary states, to solve the Gor'kov equations and to obtain gap equations. In the case of nonunitary state, time reversal symmetry is broken spontaneously. We are interested in the several physical properties of the nonunitary states, as recently discussed in a nonrelativistic theory~[21]. 

The investigation of the collective modes is also interesting in our theory. We have introduced new mean fields for pairings, and therefore they may have collective modes which have not been known yet. For example, various collective modes in the BW and ABM states were studied in detail in nonrelativistic theory~[6,8,22$\sim$24]. It is interesting to study how these modes are modified under the relativistic effects. 

Extension of our theory to the Eliashberg formalism ( strong coupling theory )~[25] is an important theme. We should study how the retardation of the pairing interaction modifies our results. In the context of the relativistic theory, inclusion of the Coulomb interaction ( and/or the photon-mediated interaction ) is also interesting. 

We have obtained the Green's functions of the relativistic superconductivity in this paper, with which we gain the abilities to investigate various physical properties. The problems given above may be treated by our Green's functions. These investigations demand us of preparation of the next stage, which is outside of this paper.

\acknowledgements

The author would like to thank Profs. H. Akai, K. Higashijima, Y. Hosotani, K. Ishikawa, Y. Nambu, H. Toki and K. Yamaguchi for many valuable suggestions.

\appendix

\section{Derivation of the specific heat}

We write down the thermodynamic potential ( in unit volume ) in our theory:
\begin{eqnarray}
\Omega_{s}(T) &=& \Omega_{n}(T=0)+E_{c}(T=0) \nonumber \\
& & -k_{B}T{\rm tr}\int\frac{d^{3}\bmk}{(2\pi)^{3}}\ln(1+e^{-\beta E_{+}})(1+e^{-\beta E_{-}}),
\end{eqnarray}
where the first term gives the normal state thermodynamic potential at $T=0$, the second term gives the condensation energy at $T=0$, and the third term gives the contribution of the Bogoliubov quasiparticles. The entropy is given by
\begin{eqnarray} 
S &=& -\frac{\partial \Omega_{s}(T)}{\partial T} \nonumber \\
&=& -k_{B}{\rm tr}\int\frac{d^{3}\bmk}{(2\pi)^{3}}\Bigl\{(1-f(E_{+}))\ln(1-f(E_{+}))+f(E_{+})\ln f(E_{+}) \nonumber \\
& & \qquad +(1-f(E_{-}))\ln(1-f(E_{-}))+f(E_{-})\ln f(E_{-})\Bigr\}.
\end{eqnarray}
This is the formula for the entropy of quasiparticle ideal gas. Here, $f(E_{\pm})$ is the Fermi distribution function for the Bogoliubov quasiparticles:
\begin{eqnarray}
f(E_{\pm}) &=& \frac{1}{\exp(\beta E_{\pm})+1}.
\end{eqnarray}
At $\Delta\to 0$, Eq. (A2) gives the entropy for the normal state. This indicates a second-order phase transition. We yield the formula for the specific heat in the relativistic theory:
\begin{eqnarray}
C &=& T\frac{dS}{dT} = -\beta\frac{dS}{d\beta}\nonumber \\
&=& {\rm tr}\int\frac{d^{3}\bmk}{(2\pi)^{3}}\frac{k_{B}}{2}\beta^{2}\Bigl\{(E_{+}+\beta\frac{dE_{+}}{d\beta})E_{+}f(E_{+})(1-f(E_{+}))\Bigr\} \nonumber \\
& & \qquad +(E_{-}+\beta\frac{dE_{-}}{d\beta})E_{-}f(E_{-})(1-f(E_{-})) \nonumber \\
&=& \int\frac{d^{3}\bmk}{(2\pi)^{3}}k_{B}\frac{\beta^{2}}{2}\Bigl\{(E_{+}+\beta\frac{dE_{+}}{d\beta})E_{+}\sech^{2}\frac{\beta}{2}E_{+} \nonumber \\
& & \qquad +(E_{-}+\beta\frac{dE_{-}}{d\beta})E_{-}\sech^{2}\frac{\beta}{2}E_{-}\Bigr\}.
\end{eqnarray}
Futhermore, at $T\ll T_{c}$, the gap function satisfies the condition $\Delta(T)\approx\Delta(T=0)$, and therefore $dE_{\pm}/d\beta=0$. We intend to study the temperature dependence of the heat capacity at $T\ll T_{c}$. Hence, we use a formula which neglects the temperature dependence of the gap function. Therefore, we obtain (19) in Sec. IV.

\section{Response Function for the Meissner Effect}

Our starting point is the Gor'kov equation under the presence of an external field, (13) of part I~[16,26,27]. In this equation, we regard $A_{\mu}$ as a perturbation, and see the variation in the one-particle propagator in the first order. We decompose Green's functions as $S_{F}=S^{(0)}_{F}+S^{(1)}_{F}$, $F=F^{(0)}+F^{(1)}$, $\bar{F}=\bar{F}^{(0)}+\bar{F}^{(1)}$, and put them into the Gor'kov equation, retaining only the first order terms. Here $(0)$ indicates the 0th order, the part which satisfies the Gor'kov equation with no external field, while $(1)$ indicates the variation with respect to the external field. Due to the choice of the Coulomb gauge, we assume that the mean field is rigid under the external field~[26]. We obtain 
\[
\left(
\begin{array}{cccc}
i\gamma^{\mu}\partial_{\mu}-m+\gamma^{0}\mu & \Delta \\
\bar{\Delta} & i\gamma^{\mu T}\partial_{\mu}+m-\gamma^{0T}\mu 
\end{array}
\right)
\left(
\begin{array}{cccc}   
S^{(1)}_{F}(x,y) & -iF^{(1)}(x,y) \\
-i\bar{F}^{(1)}(x,y) & -S^{(1)}_{F}(y,x)^{T}
\end{array}
\right)
\]
\begin{equation}
=
\left(
\begin{array}{cccc}
-e\gamma^{\mu}A_{\mu}(x) & 0 \\
0 & e\gamma^{\mu T}A_{\mu}(x)   
\end{array}
\right)
\left(
\begin{array}{cccc}
S^{(0)}_{F}(x,y) & -iF^{(0)}(x,y) \\
-i\bar{F}^{(0)}(x,y) & -S^{(0)}_{F}(y,x)^{T} 
\end{array}
\right),
\end{equation}
and write it in the form of the Dyson-type integral equation, 
\[
\left(
\begin{array}{cccc}
S^{(1)}_{F}(x,y) & -iF^{(1)}(x,y) \\
-i\bar{F}^{(1)}(x,y) & -S^{(1)}_{F}(y,x)^{T} 
\end{array}
\right)
=\int d^{4}z 
\]
\begin{equation}
\times
\left(
\begin{array}{cccc}
S^{(0)}_{F}(x,z) & -iF^{(0)}(x,z) \\
-i\bar{F}^{(0)}(x,z) & -S^{(0)}_{F}(z,x)^{T} 
\end{array}
\right)
\left(
\begin{array}{cccc}
-e\gamma^{\mu}A_{\mu}(z) & 0 \\
0 & e\gamma^{\mu T}A_{\mu}(z) 
\end{array}
\right)
\left(
\begin{array}{cccc}
S^{(0)}_{F}(z,y) & -iF^{(0)}(z,y) \\
-i\bar{F}^{(0)}(z,y) & -S^{(0)}_{F}(y,z)^{T} 
\end{array}
\right).
\end{equation}
We take the first order variation, for example for $S_{F}$:
\begin{eqnarray}
S^{(1)}_{F}(x,y) &=& -e\int d^{4}z \Bigl( 
S^{(0)}_{F}(x,z)\gamma^{\mu}A_{\mu}(z)S^{(0)}_{F}(z,y)
+ F^{(0)}(x,z)\gamma^{\mu T}A_{\mu}(z)\bar{F}^{(0)}(z,y) \Bigr).  \nonumber \\
& &   
\end{eqnarray}
The induced current now becomes  
\begin{eqnarray}
j^{\mu}(x) &=& ie \lim_{y\to x^{+}} {\rm tr}\gamma^{\mu}S^{(1)}_{F}(x,y) \nonumber \\
&=& -ie^{2}\int d^{4}z {\rm tr}\Bigl( \gamma^{\mu}S^{(0)}_{F}(x,z)\gamma^{\nu}S^{(0)}_{F}(z,x) + \gamma^{\mu}F^{(0)}(x,z)\gamma^{\mu T}\bar{F}^{(0)}(z,x) \Bigr)A_{\nu}(z) \nonumber \\
&=&  e^{2}\int d^{4}z \Pi_{\mu\nu}(x,z)A^{\nu}(z). 
\end{eqnarray}
We obtain the polarization function as
\begin{eqnarray}
\Pi_{\mu\nu}(x,z) &=& -ie^{2}{\rm tr}\Bigl( \gamma^{\mu}S^{(0)}_{F}(x,z)\gamma^{\nu}S^{(0)}_{F}(z,x) + \gamma^{\mu}F^{(0)}(x,z)\gamma^{\nu T}\bar{F}^{(0)}(z,x) \Bigr).
\end{eqnarray}
This is given in the Fourier transform as
\begin{eqnarray}
\Pi_{\mu\nu}(q) &=& -ie^{2}{\rm tr}\int\frac{d^{4}k}{(2\pi)^{4}}\Bigl( \gamma^{\mu}S^{(0)}_{F}(k)\gamma^{\nu}S^{(0)}_{F}(k+q) + \gamma^{\mu}F^{(0)}(k)\gamma^{\nu T}\bar{F}^{(0)}(k+q) \Bigr). \nonumber \\
& & 
\end{eqnarray}
Following the same way, we obtain formulae in the Matsubara formalism. The induced current becomes 
\begin{eqnarray}
j^{\mu}(x) &=& -e \lim_{y\to x^{+}}{\rm tr}\gamma^{\mu}{\cal S}^{(1)}(x,y). 
\end{eqnarray}
The polarization becomes
\begin{eqnarray}
\Pi_{\mu\nu}(x,z) &=& e^{2}{\rm tr}\Bigl(\gamma^{\mu}{\cal S}^{(0)}(x,z)\gamma^{\nu}{\cal S}^{(0)}(z,x)-\gamma^{\mu}{\cal F}^{(0)}(x,z)\gamma^{\nu T}\bar{{\cal F}}^{(0)}(z,x)\Bigl),
\end{eqnarray}
and in the Fourier transform
\begin{eqnarray}
\Pi_{\mu\nu}(q) &=& e^{2}\sum_{n}\frac{1}{\beta}\int\frac{d^{3}\bmk}{(2\pi)^{3}}{\rm tr}\Bigl(\gamma^{\mu}{\cal S}^{(0)}(k)\gamma^{\nu}{\cal S}^{(0)}(k+q)-\gamma^{\mu}{\cal F}^{(0)}(k)\gamma^{\nu T}\bar{{\cal F}}^{(0)}(k+q)\Bigr). \nonumber \\
& &          
\end{eqnarray}
Hereafter we use only the Matsubara formalism. 

In the gauge invariant theory, the transversal condition for the polarization, $q^{\mu}\Pi_{\mu\nu}=0$, comes from the current conservation. In the case of zero temperature quantum field theory for the vacuum, from the Lorentz invariance, it takes the form $\Pi_{\mu\nu}(q)=(g_{\mu\nu}q^{2}-q_{\mu}q_{\nu})\Pi(q^{2})$, which represents the 4-dimensinal transversality. In our case, the system is under finite-temperature and finite-density. We take a specific coordinate for the imaginary time and select a suitable rest frame for the description of the system. Then in our case, the Lorentz invariance is lost. Only $O(3)$ rotational invariance in $\bmk$-space remains. Then we use the following projection operator ( here $q=(q_{0}=i\frac{2n+1}{\beta}\pi=i\omega_{n},\bmq)$ )~[28]: 
\begin{eqnarray}
\Pi_{\mu\nu}(q) &=& P^{(L)}_{\mu\nu}(q)\Pi^{(L)}(q_{0},|\bmq|) + P^{(T)}_{\mu\nu}(q)\Pi^{(T)}(q_{0},|\bmq|),  
\end{eqnarray}
and from $P^{(T)}_{\mu\nu}+P^{(L)}_{\mu\nu}=g_{\mu\nu}-q_{\mu}q_{\nu}/q^{2}$, we derive a 3-dimensional transversal condition, and then
\begin{equation}
P^{(L)}_{\mu\nu}(q) = -\frac{1}{q^{2}\bmq^{2}}\left(
\begin{array}{cccc}
(\bmq^{2})^{2} & \bmq^{2}q_{0}q_{j} \\
\bmq^{2}q_{i}q_{0} & (q_{0})^{2}q_{i}q_{j} 
\end{array}
\right),
\end{equation}
\begin{equation}
P^{(T)}_{\mu\nu}(q) = \frac{1}{\bmq^{2}}\left(
\begin{array}{cccc}
0 & 0 \\
0 & g_{ij}\bmq^{2}+q_{i}q_{j} 
\end{array}
\right).
\end{equation}
$P^{(T)}_{\mu\nu}$ is for the 3-dimensional transversal condition, while $P^{(L)}_{\mu\nu}$ is the 3-dimensional longitudinal condition, and each of them satisfies $q^{\mu}\Pi^{(T,L)}_{\mu\nu}=0$. Using them, we obtain
\begin{eqnarray}
\Pi^{(L)}(q_{0},|\bmq|) &=& -\frac{q^{2}}{\bmq^{2}}\Pi_{00}(q), \\
\Pi^{(T)}(q_{0},|\bmq|) &=& \frac{1}{2}\Bigl(\Pi^{\mu}_{ \mu}(q)+\frac{q^{2}}{\bmq^{2}}\Pi_{00}(q)\Bigr). 
\end{eqnarray}
Next we choose the Coulomb gauge $q^{i}A_{i}(q)=0$ and also take $A_{0}=0$, and the response becomes only 3-dimensinal transversal:
\begin{eqnarray}
j_{i}(q) &=& -\Pi^{(T)}(q)A_{i}(q).
\end{eqnarray}
For the Meissner effect, we only consider the static field $q_{0}=0$, then 
\begin{eqnarray}
{\bf j}(0,\bmq) &=& -\frac{1}{2}(\Pi_{11}+\Pi_{22}+\Pi_{33})(0,\bmq){\bf A}(0, \bmq) \nonumber \\
&\equiv& -\Pi(0, \bmq){\bf A}(0,\bmq). 
\end{eqnarray}
We calculate this response by our Green's functions. We concentrate on the case of the scalar $\Delta^{S}$ and the 0th component of the vector $\Delta^{V}_{0}$. For the finite $\bmq$, it becomes rather complicated calculation. Here we only want to study the Meissner effect, then we take the London limit $\bmq\to 0$. After the frequency summation, we obtain
\begin{eqnarray}
\lim_{\bmq\to 0}\Pi(0,\bmq) &=& -e^{2}\int\frac{d^{3}\bmp}{(2\pi)^{3}}\frac{1}{(E^{2}_{+}(\bmp)-E^{2}_{-}(\bmp))^{2}}\Big\{ \nonumber \\
& & \frac{1}{E_{+}(\bmp)^{2}}N(E_{+}(\bmp),\bmp)\frac{\beta}{2}\sech^{2}\frac{\beta}{2}E_{+}(\bmp) \nonumber \\
&+&\frac{1}{E_{-}(\bmp)^2}N(E_{-}(\bmp),\bmp)\frac{\beta}{2}\sech^{2}\frac{\beta}{2}E_{-}(\bmp) \nonumber \\
&-&\frac{1}{E_{+}(\bmp)^{3}}\frac{5E_{+}(\bmp)^{2}-E_{-}(\bmp)^{2}}{E_{+}(\bmp)^{2}-E_{-}(\bmp)^{2}}N(E_{+}(\bmp),\bmp)\tanh\frac{\beta}{2}E_{+}(\bmp) \nonumber \\
&-&\frac{1}{E_{-}(\bmp)^{3}}\frac{E_{+}(\bmp)^{2}-5E_{-}(\bmp)^{2}}{E_{+}(\bmp)^{2}-E_{-}(\bmp)^{2}}N(E_{-}(\bmp),\bmp)\tanh\frac{\beta}{2}E_{-}(\bmp) \Big\},
\end{eqnarray}
where $N(p_{0},\bmp)$ for the scalar $\Delta^{S}$ case is  
\begin{eqnarray}
N(p_{0},\bmp) &=& 2\Big\{3(p^{4}_{0}-\mu^{4})(p^{2}_{0}-\mu^{2})-6(p^{2}_{0}-\mu^{2})^{2}(\bmp^{2}+m^{2}) \nonumber \\
& & +(p^{2}_{0}+\mu^{2})(\bmp^{2}+m^{2})(5\bmp^{2}+9m^{2}) \nonumber \\
& & -(p^{4}_{0}+6\mu^{2}p^{2}_{0}+\mu^{4}+(\bmp^{2}+m^{2})^{2})(\bmp^{2}+3m^{2})\nonumber \\
& & +|\Delta^{S}|^{2}[-(p^{2}_{0}-\mu^{2})(3p^{2}_{0}+9\mu^{2}-2\bmp^{2}-6m^{2})+(\bmp^{2}+m^{2})(\bmp^{2}-3m^{2})] \nonumber \\
& & +|\Delta^{S}|^{4}[-3(p^{2}_{0}-3\mu^{2})+5\bmp^{2}+3m^{2}]+3|\Delta^{S}|^{6}\Big\}, 
\end{eqnarray}
and for the vector $\Delta^{V}_{0}$ case is 
\begin{eqnarray}
N(p_{0},\bmp) &=& 2\Big\{3(p^{4}_{0}-\mu^{4})(p^{2}_{0}-\mu^{2})-6(p^{2}_{0}-\mu^{2})^{2}(\bmp^{2}+m^{2}) \nonumber \\
& & +(p^{2}_{0}+\mu^{2})(\bmp^{2}+m^{2})(5\bmp^{2}+9m^{2}) \nonumber \\
& & -(p^{4}_{0}+6\mu^{2}p^{2}_{0}+\mu^{4}+(\bmp^{2}+m^{2})^{2})(\bmp^{2}+3m^{2}) \nonumber \\
& & -|\Delta^{V}_{0}|^{2}[3(3p^{4}_{0}-2\mu^{2}p^{2}_{0}-\mu^{4})-(3p^{2}_{0}-\mu^{2})(\bmp^{2}+3m^{2}) \nonumber \\
& & +(p^{2}_{0}+\mu^{2})(\bmp^{2}-9m^{2})-3(\bmp^{2}+m^{2})(\bmp^{2}-3m^{2})+4\bmp^{2}(\bmp^{2}-m^{2})] \nonumber \\
& & +|\Delta^{V}_{0}|^{4}[3(3p^{2}_{0}-\mu^{2})+5\bmp^{2}-9m^{2}]-3|\Delta^{V}_{0}|^{6}\Big\}.
\end{eqnarray}
In (B17), the first and second terms become zero at $T=0$, and only contribute the neighbor of the Fermi surface to the integral. The integral of these two terms always converge. These terms have their origin on the terms which describe collisionless elementary excitations~[29]. The third and fourth terms give diverged integrals, and we interpret these terms as coming from the terms which describe the creation-annihilation of the Cooper pair~[29]. The third and fourth terms have small temperature dependences. In the finite-temperature field theory, $\Pi_{\mu\nu}$ is devided into $\Pi_{\mu\nu}=\Pi^{matter}_{\mu\nu}+\Pi^{vacuum}_{\mu\nu}$ with the definition $\Pi^{vacuum}_{\mu\nu}\equiv\lim_{\mu\to 0,T\to 0}\Pi_{\mu\nu}$, and the usual renormalization procedure is adopted to the $\Pi^{vacuum}_{\mu\nu}$~[28,30]. Here we also use this method to handle the divergence of (B17). Therefore, 
\begin{eqnarray}
\Pi_{\mu\nu}(q) &=& \{\Pi_{\mu\nu}(q)-\Pi^{vacuum}_{\mu\nu}(q)\}+\Pi^{vacuum}_{\mu\nu}(q),
\end{eqnarray}
and
\begin{eqnarray}
\Pi^{vacuum}_{\mu\nu}(q) &=& -ie^{2}\int\frac{d^{4}p}{(2\pi)^{4}}{\rm tr}\gamma_{\mu}\frac{\kfey+m}{k^{2}-m^{2}}\gamma_{\nu}\frac{\kfey+\qfey+m}{(k+q)^{2}-m^{2}}.
\end{eqnarray}
It is well-known that, after the method of usual gauge-invariant regularization and renormalization is performed to $\Pi^{vacuum}_{\mu\nu}(q)$, this term becomes $(g_{\mu\nu}q^{2}-q_{\mu}q_{\nu})\Pi(q^{2})$, and vanishes at $q_{0}=0, \bmq\to 0$. Therefore, for our aim, $\lim_{\bmq\to 0}\{\Pi_{\mu\nu}(0,\bmq)-\Pi^{vacuum}_{\mu\nu}(0,\bmq)\}$ should be calculated. By using this function, we examine (B16) at $\bmq\to 0$. This examination corresponds to treat the London equation at $\bmq\to 0$. Thus, the temperature dependence of the integral (B17) corresponds to the temperature dependence of the supercurrent in the relativistic superconductivity.

\begin{figure}

\caption{A schematic figure of the excitation energy spectra of the quasiparticles of the $\Delta^{S}$ and $\Delta^{V}_{0}$ states. $E_{+}$ and $E_{-}$ correspond to the quasiparticles coming from the positive energy states and the negative energy states, respectively. In $E_{+}$, the minimum energy appears at a momentum denoted by $k_{min}$, which is different between the $\Delta^{S}$ case and $\Delta^{V}_{0}$ case, which are explicitly written below Eq. (2) in the text.}

\vspace{10mm}

\caption{A schematic figure of the density of states ( DOS ) for the spin singlet pairing states, $\Delta^{S}$ and $\Delta^{V}_{0}$, normalized by $\rho_{n}(\epsilon_{F})$ as a function of energy. The region around the gap is depicted. We only take into account the contributions of the quasiparticles coming from positive energy states. In the case of $\Delta^{S}$, the DOS approaches to infinity at $\omega\to\pm|\Delta^{S}|$, while in $\Delta^{V}_{0}$, the DOS becomes infinity at $\omega \to \pm m|\Delta^{V}_{0}|(\mu^{2}+|\Delta^{V}_{0}|^{2})^{-\frac{1}{2}}$.}

\vspace{10mm}

\caption{$g_{0}$ dependence of the pairing gap of spin singlet states at $T=0$. (a) The usual solid metal. We set $\mu-m=5$eV ($r_{s}=3.2$). (b) The relativistic plasma. We set $\mu-m=0.5$MeV ($r_{s}=0.0082$). (c) The neutron star. We set $\mu-m=20$MeV. The masses are 0.5 MeV for electron and 1 GeV for neutron. All calculations are performed with the cutoff $\Lambda/(\mu-m)=1$.}

\vspace{10mm}

\caption{The cutoff dependence of the pairing gap at $T=0$. (a) The usual solid metal. We set $\mu-m=5$eV and $g_{0}=9.48\times10^{-9}$eV$^{-2}$. (b) The relativistic plasma. We set $\mu-m=0.5$MeV and $g_{0}=23.6$MeV$^{-2}$. (c) The neutron star. We set $\mu-m=20$MeV and $g_{0}=1.38\times10^{-4}$MeV$^{-2}$.}

\vspace{10mm}

\caption{The temperature dependence of the pairing gap. (a) The usual solid metal. We set $\mu-m=5$eV and $g_{0}=4.42\times10^{-9}$eV$^{-2}$. (b) The metallic hydrogen. We set $\mu-m=200$eV ( $r_{s}=0.5$ ) and $g_{0}=6.90\times10^{-10}$eV$^{-2}$. (c) The relativistic plasma. We set $\mu-m=0.5$MeV and $g_{0}=19.7$MeV$^{-2}$. (d) The neutron star. We set $\mu-m=20$MeV  and $g_{0}=7.89\times10^{-5}$MeV$^{-2}$. All calculations are done by taking the cutoff $\Lambda/(\mu-m)=1$. }

\vspace{10mm}

\caption{The presentation of our results for various electron systems in an unified view point, by functions of the chemical potential. (a) The coupling constant so as to provide $|\Delta(T=0)|/(\mu-m)=10^{-3}$ for the solution of the ``nonrelativistic'' case. By using the data given in (a), we calculate (b) the magnitudes of gaps at $T=0$, and (c) $T_{c}$ in 4-cases. All of the calculations are performed by taking the cutoff $\Lambda/(\mu-m)=1$.}

\vspace{10mm}

\caption{The temperature dependences of spin triplet gap functions. (a) The usual solid metal. We set $\mu-m=5$eV and $g_{1}=1.1\times10^{-7}$eV$^{-2}$. (b) The relativistic plasma. We set $\mu-m=0.5$eV and $g_{1}=12$MeV$^{-2}$. (c) The neutron star. We set $\mu-m=80$MeV and $g_{1}=1.5\times10^{-5}$MeV$^{-2}$. All calculations are performed with the cutoff $\Lambda/(\mu-m)=1$. The states not shown in this figure give extremely small solutions or cannot be solved because of numerical difficulties.}

\end{figure}

\begin{table}
\caption{Typical order of physical quantities in various systems.  The masses are 0.5 MeV for the elctron, and 1 GeV for the nucleon.}
\begin{tabular}{cccccc}
system & particle density (cm$^{-3}$) & $\epsilon_{F}-m$ & $|\Delta(T=0)|$ & $T_{c}$ & $\frac{|\Delta(T=0)|}{\epsilon_{F}-m}$ \\
\hline
usual solid metal\tablenote{Refs. 7, 8} & $\sim10^{22}$ & $\sim5$eV & $\sim10^{-3}-10^{-4}$eV & $<100$K & $\sim10^{-3}-10^{-4}$ \\
metallic hydrogen\tablenote{Refs. 9, 10} & $\sim10^{23}-10^{25}$ & $\sim20-200$eV & $\sim10^{-1}-10^{-2}$eV & $\sim10^{2}$K & $\sim10^{-3}-10^{-4}$ \\
``$1s$ electron of uranium'' gas\tablenote{Ref.13} & $\sim10^{29}$ & $\sim0.25$MeV & {} & {} & {} \\
relativistic plasma\tablenote{Ref.11} & $\sim10^{32}$ & $\sim2.5$MeV & {} & {} & {} \\
neutron star\tablenote{Ref.12} & $\sim10^{37}-10^{38}$ & $\sim10-100$MeV & $\sim0.1-2.5$MeV & $\sim0.1-1.5$MeV & $\sim10^{-1}-10^{-2}$ \\
\end{tabular}
\end{table}

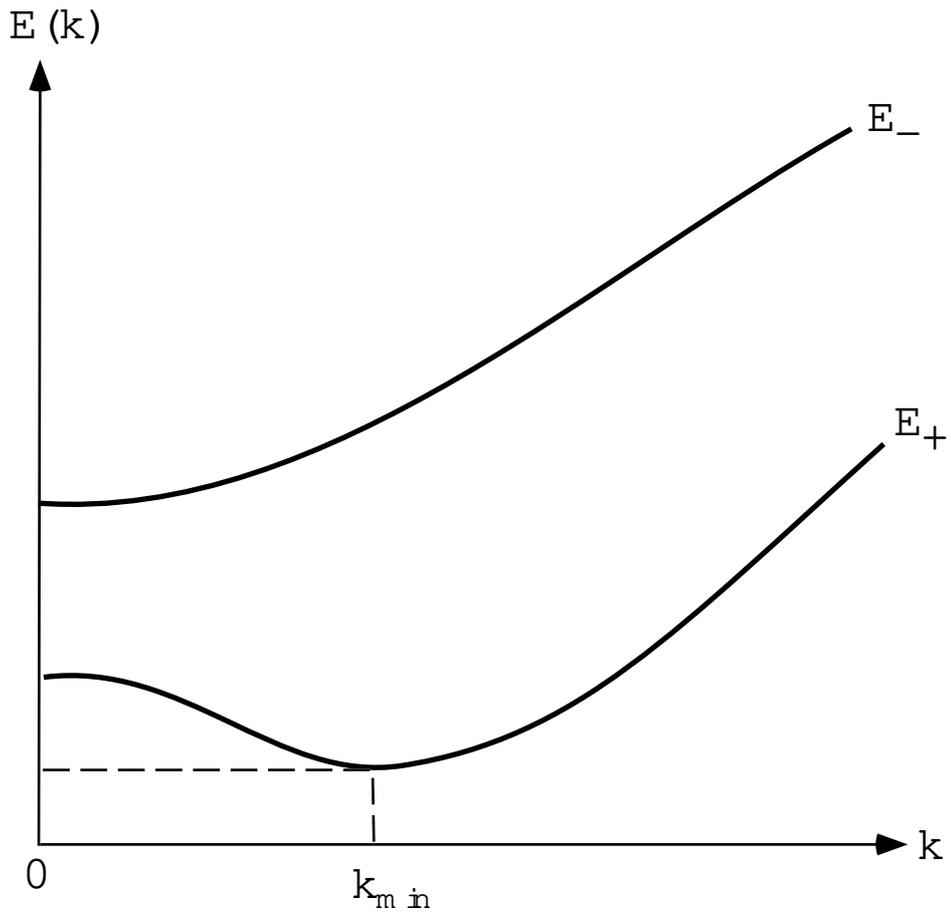
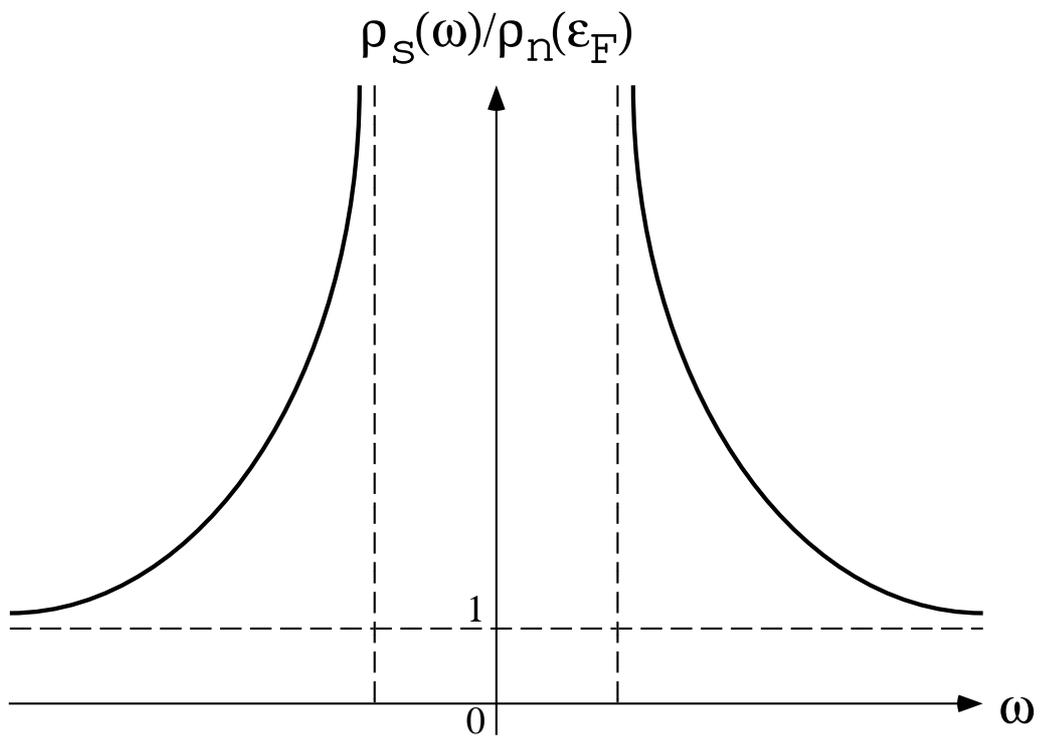
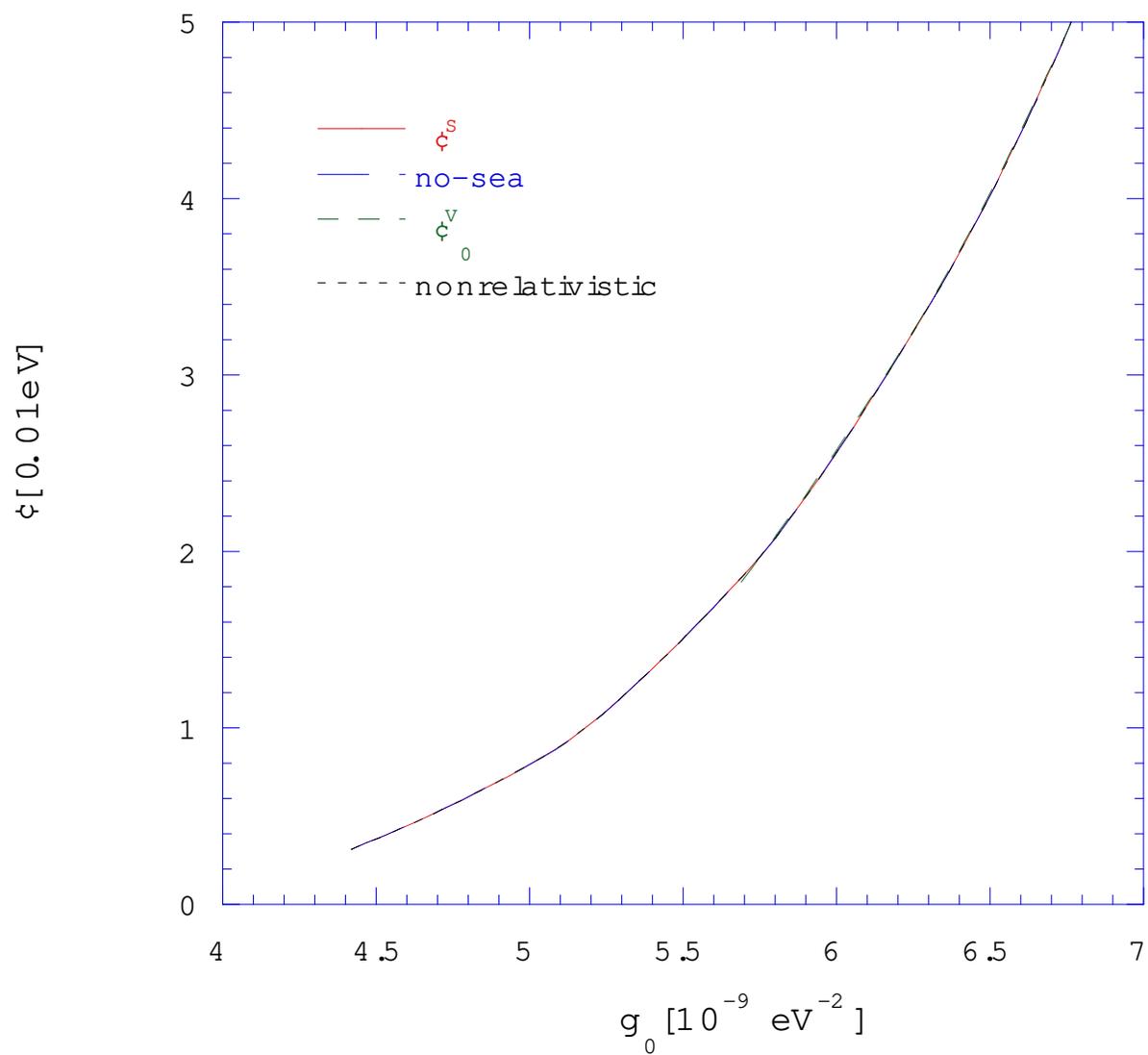
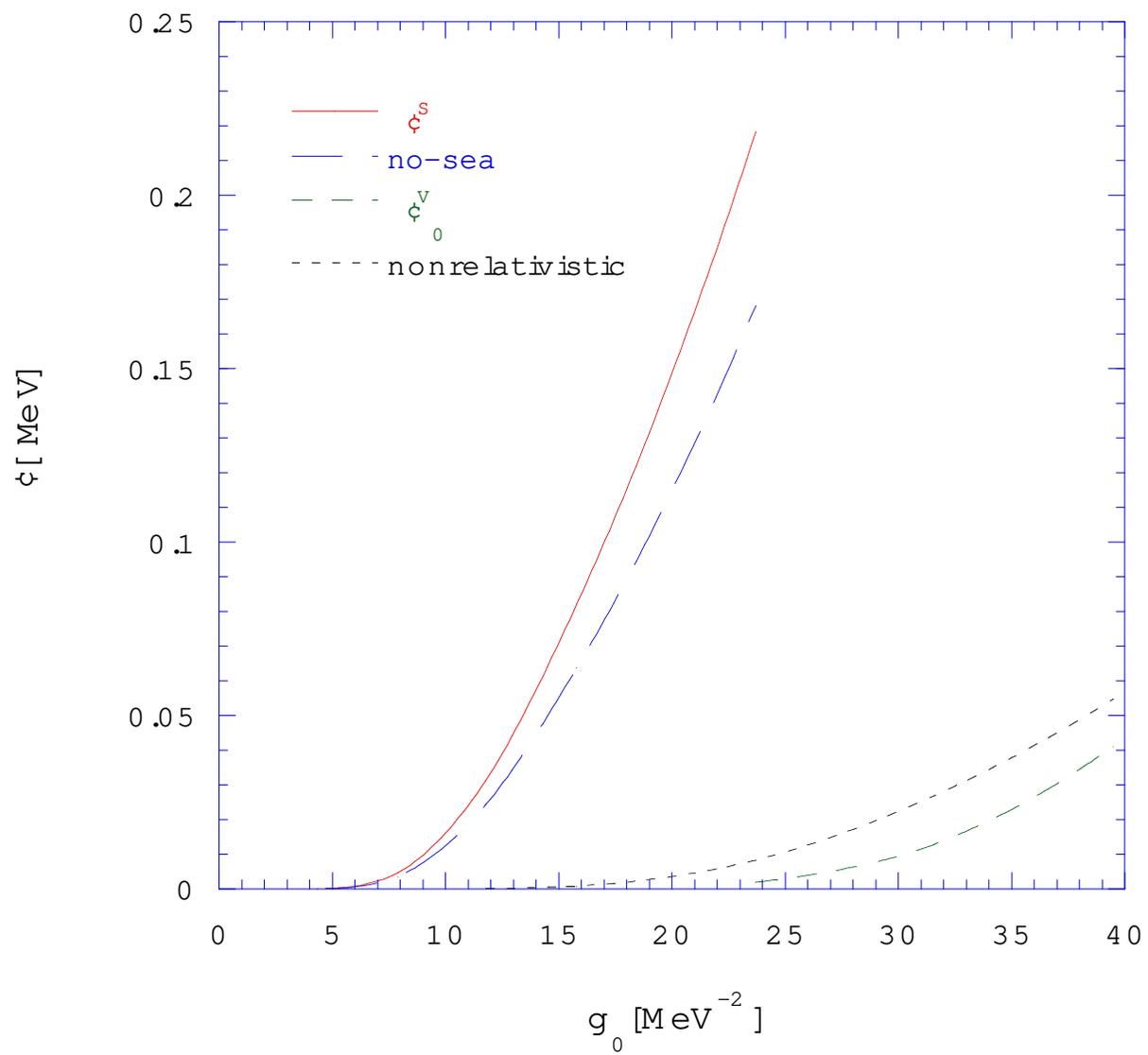
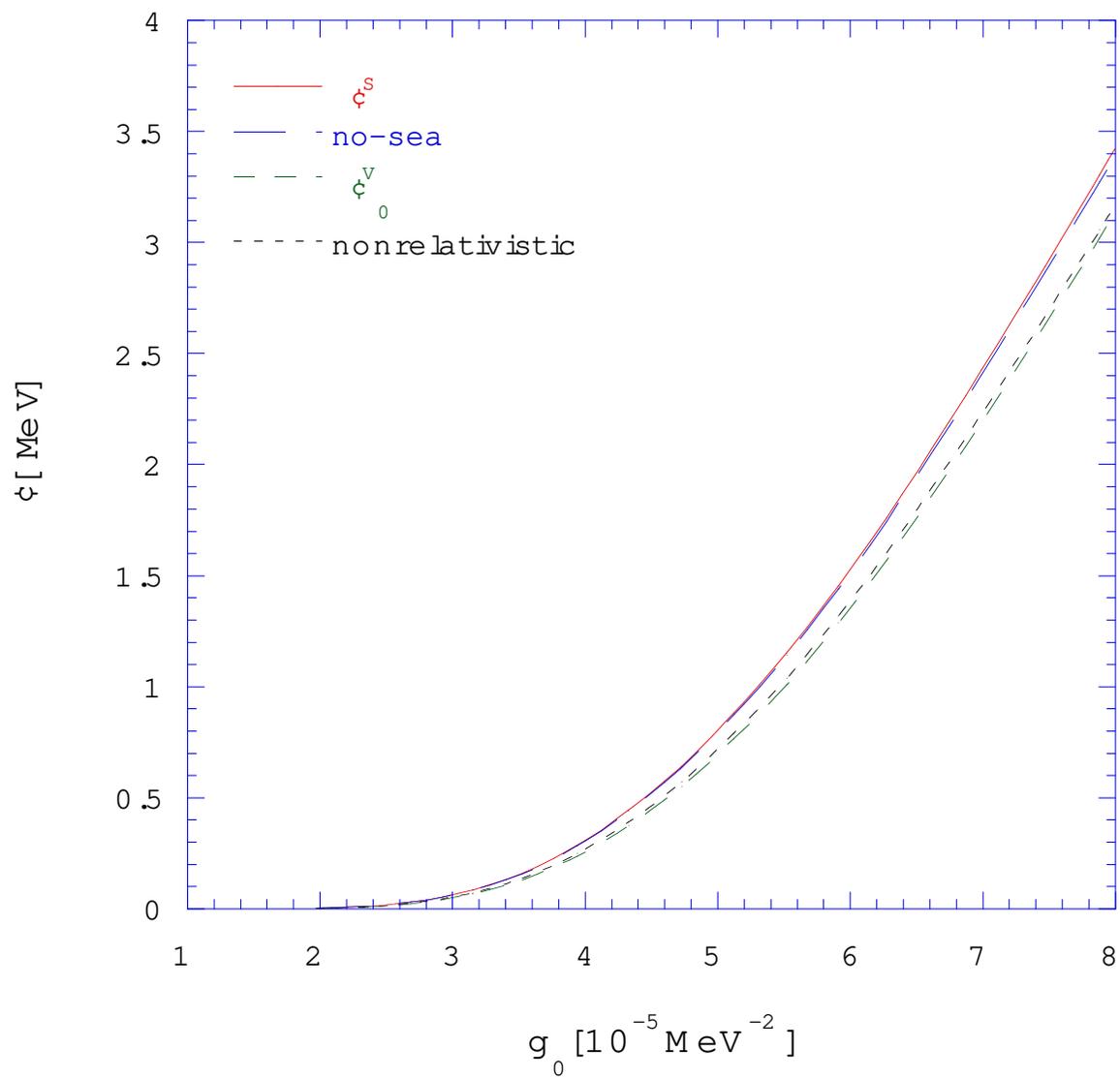
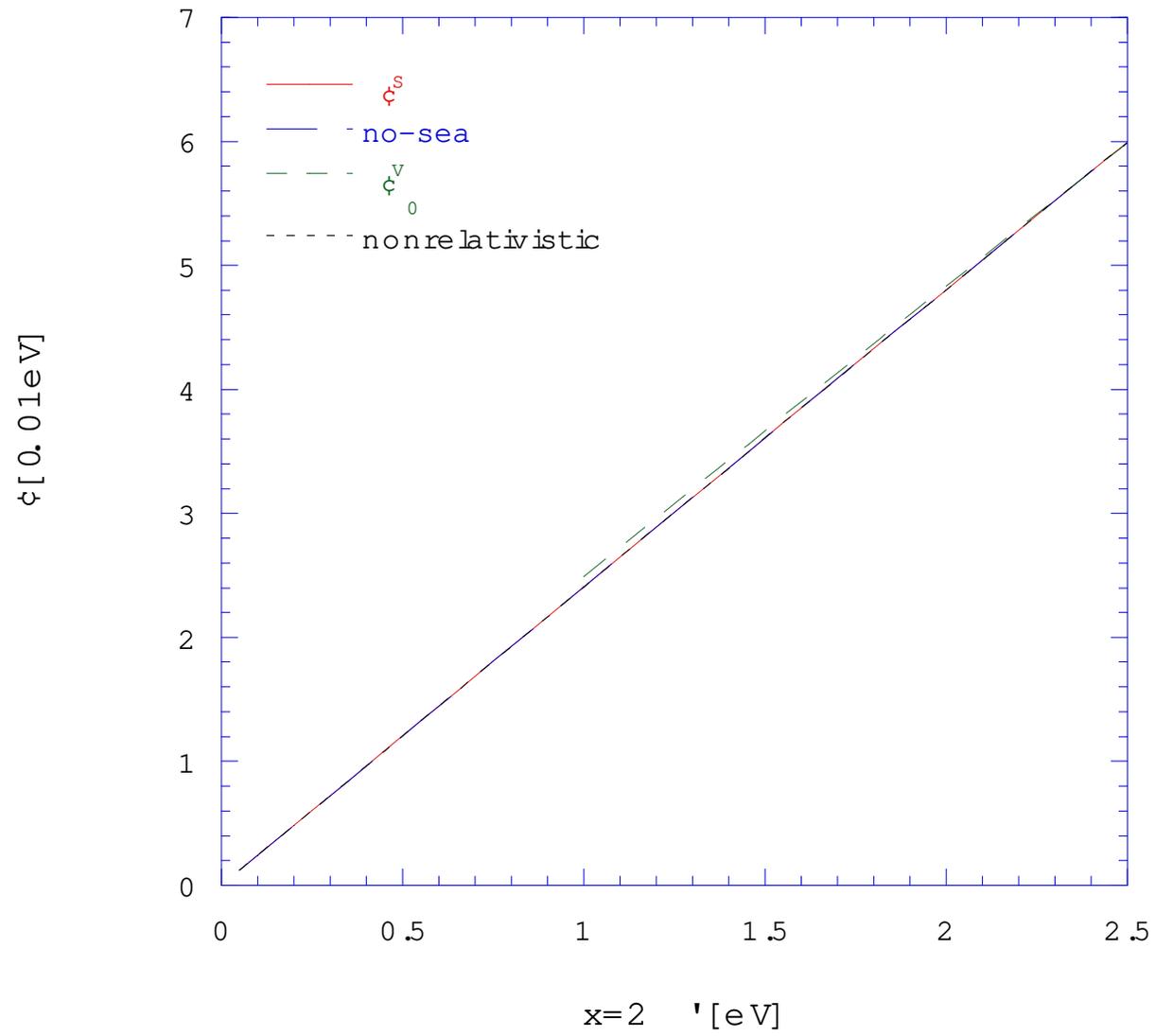
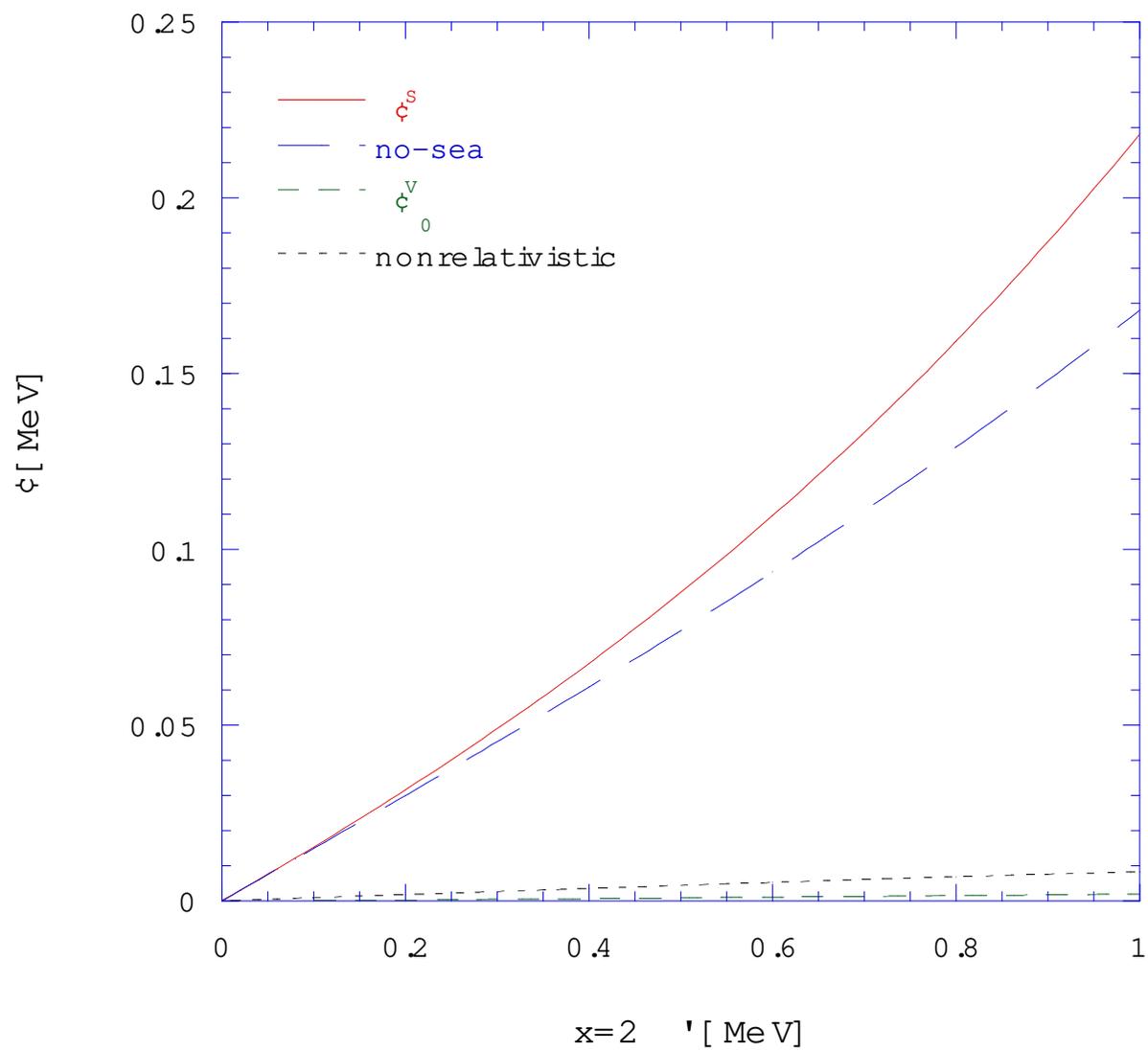
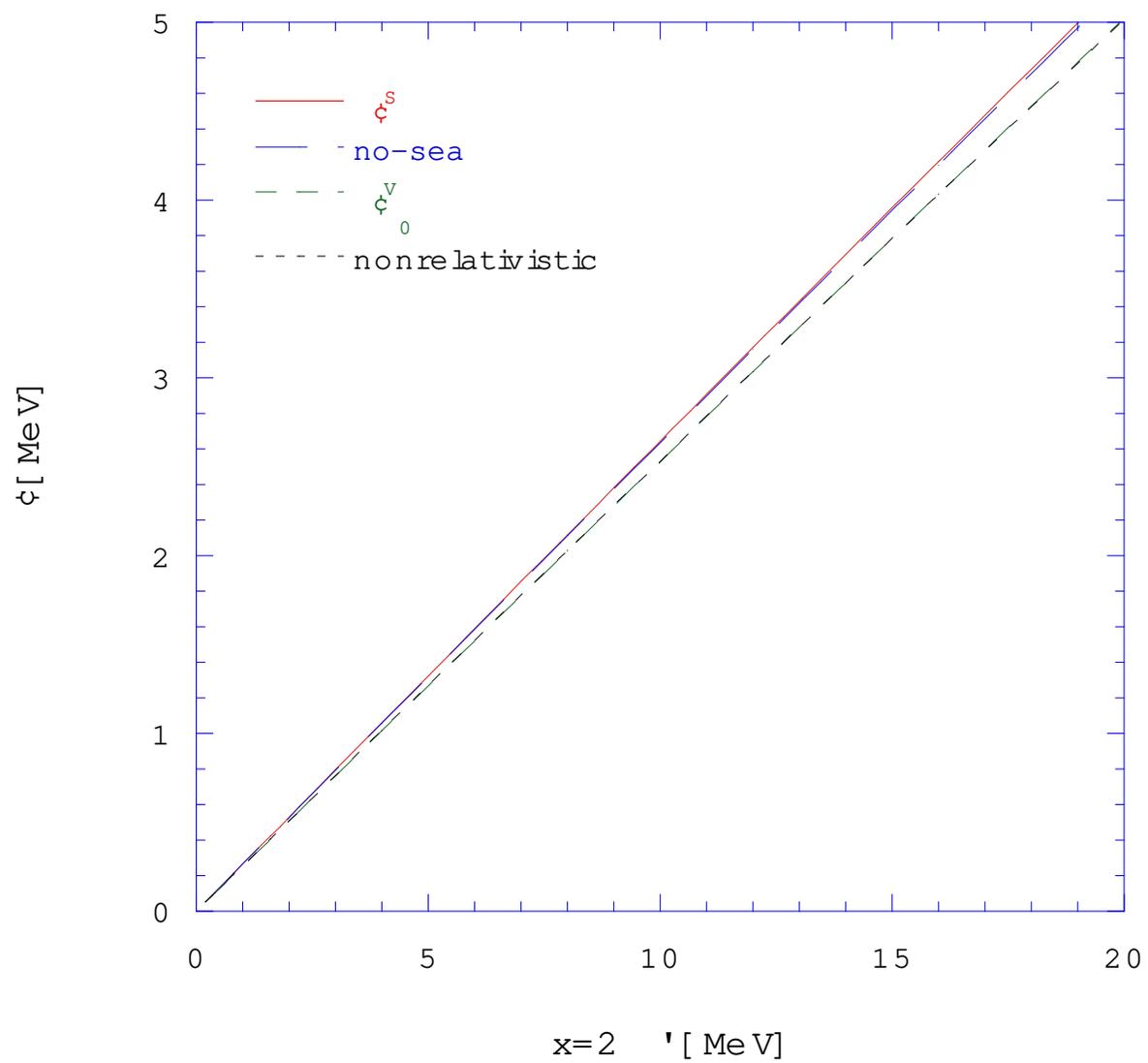
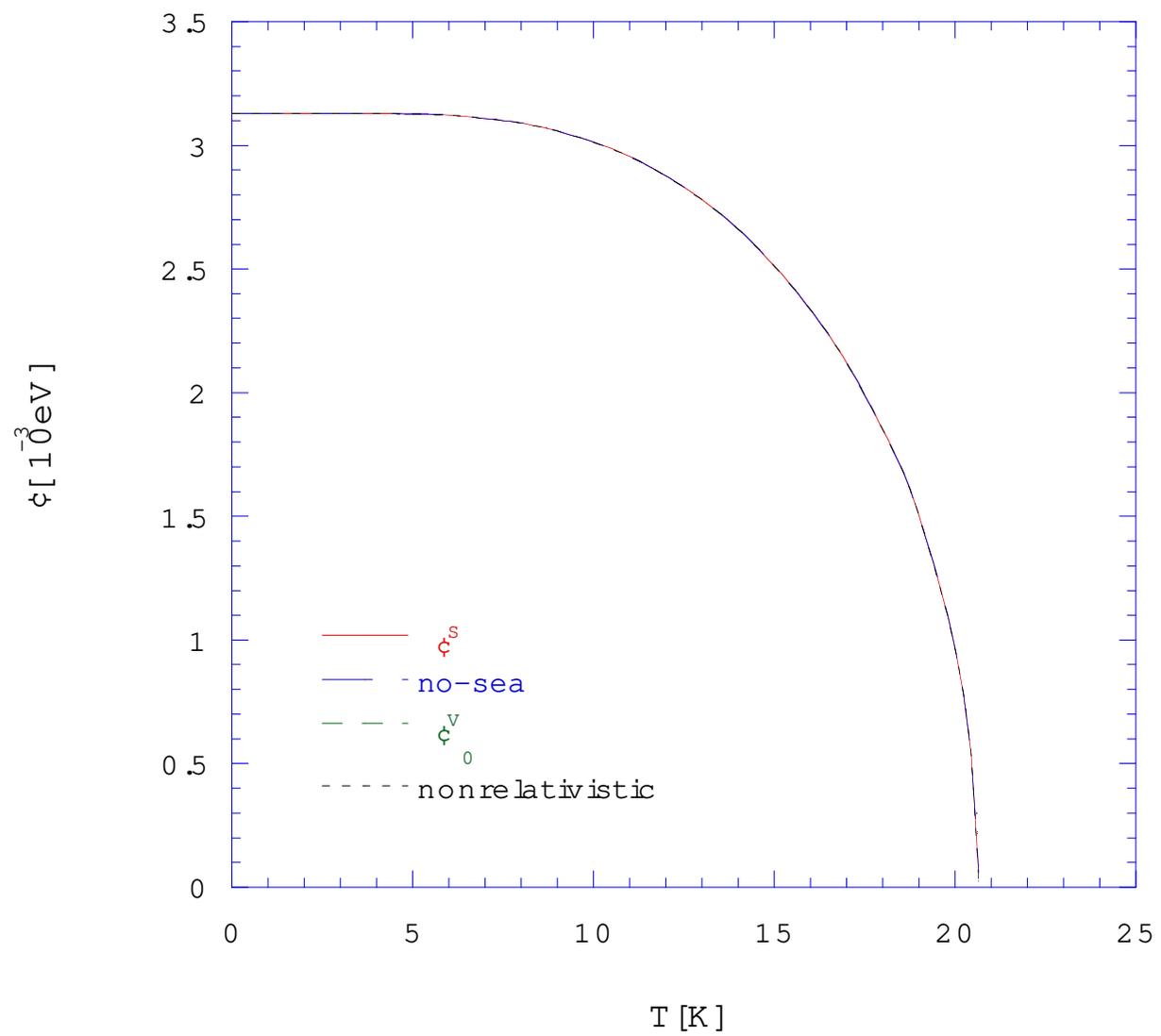
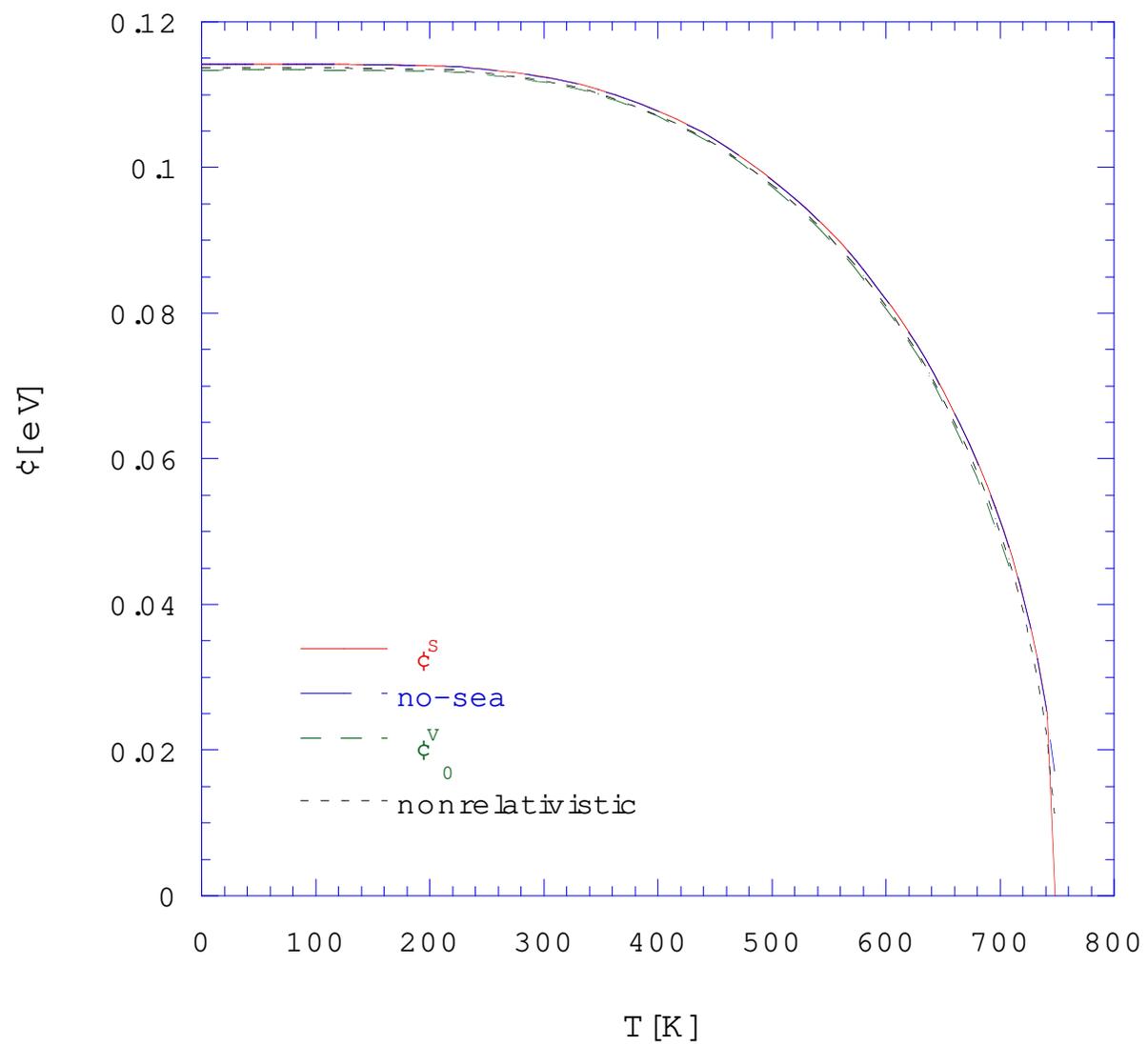
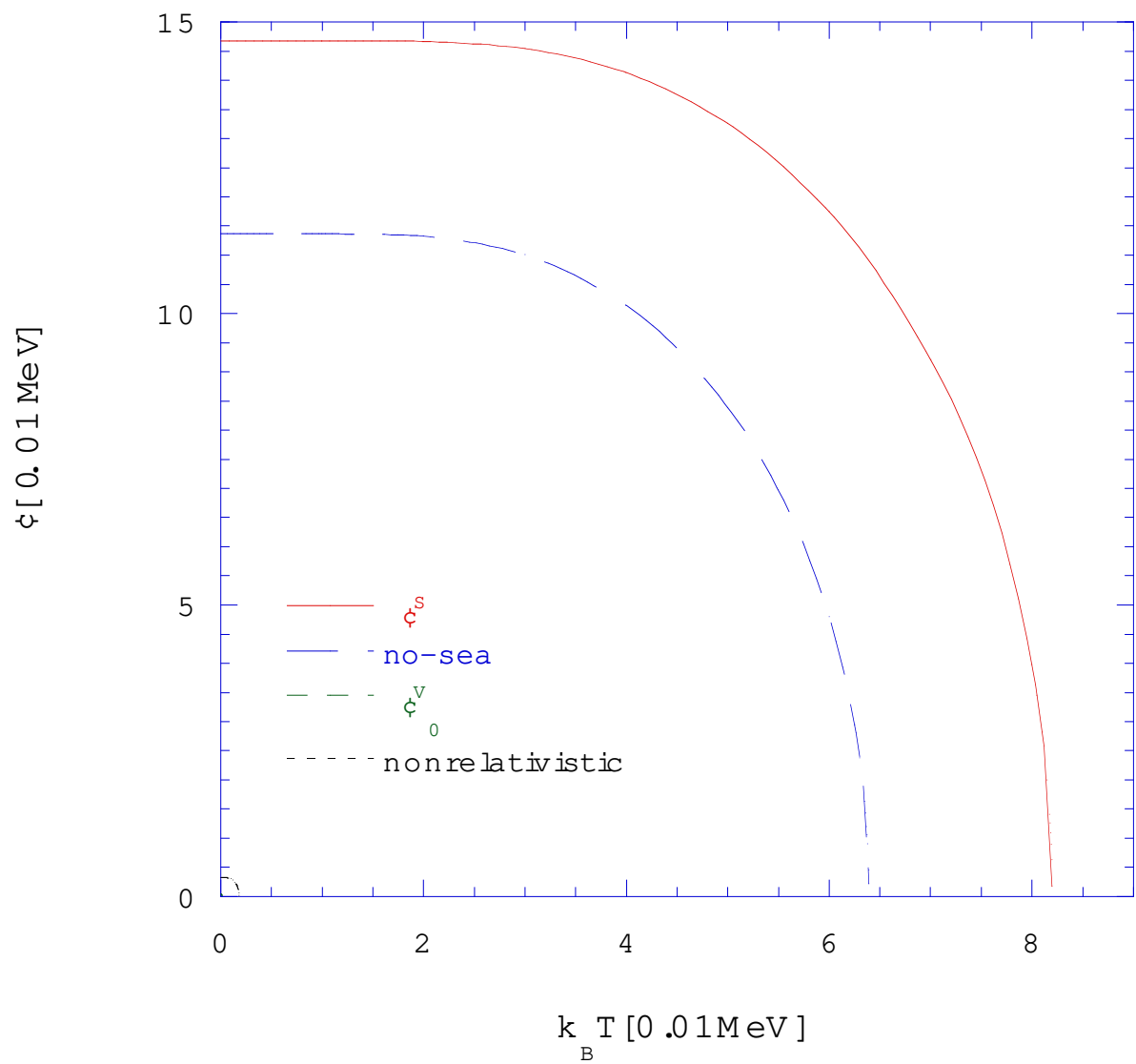
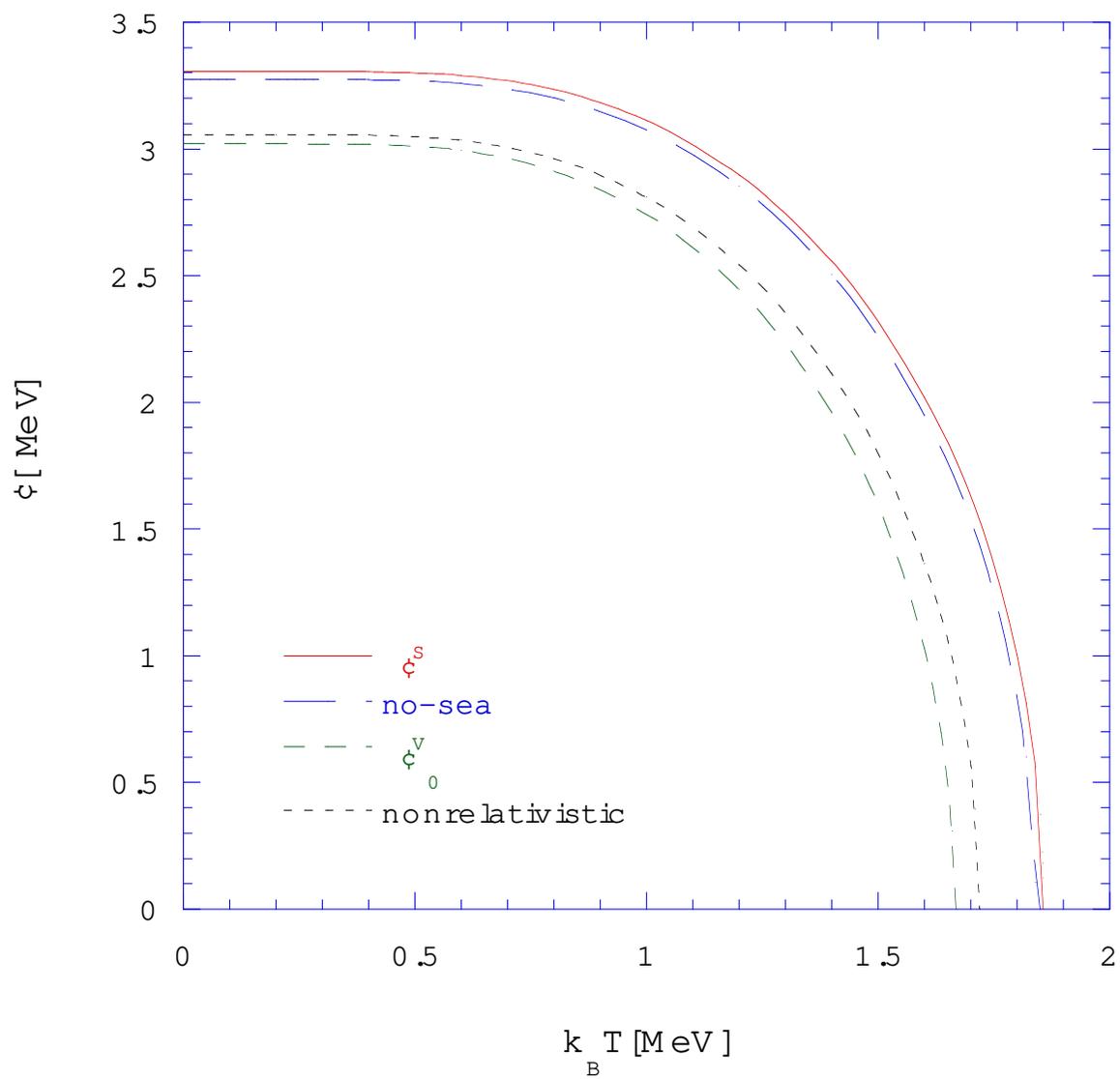
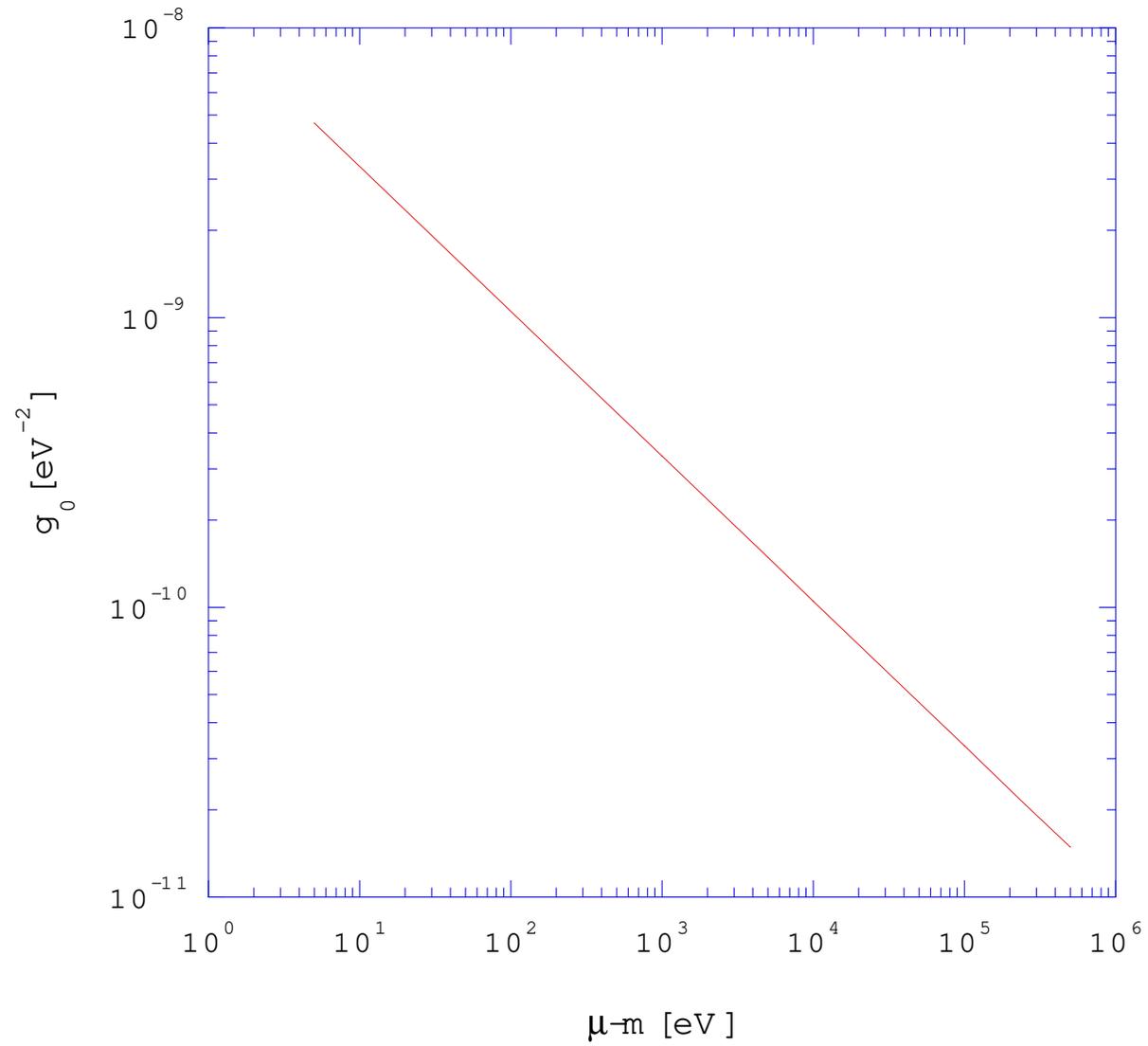
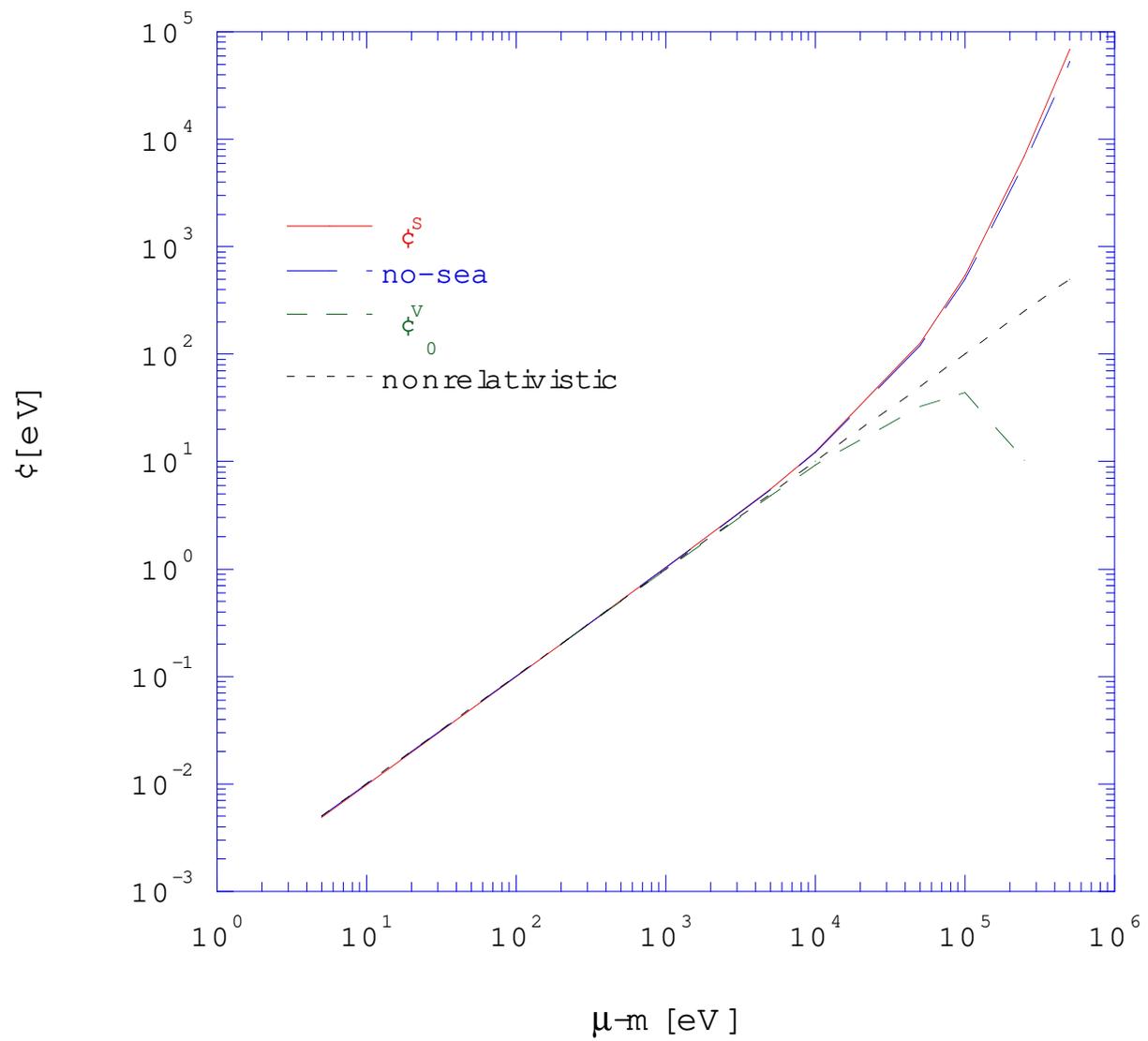
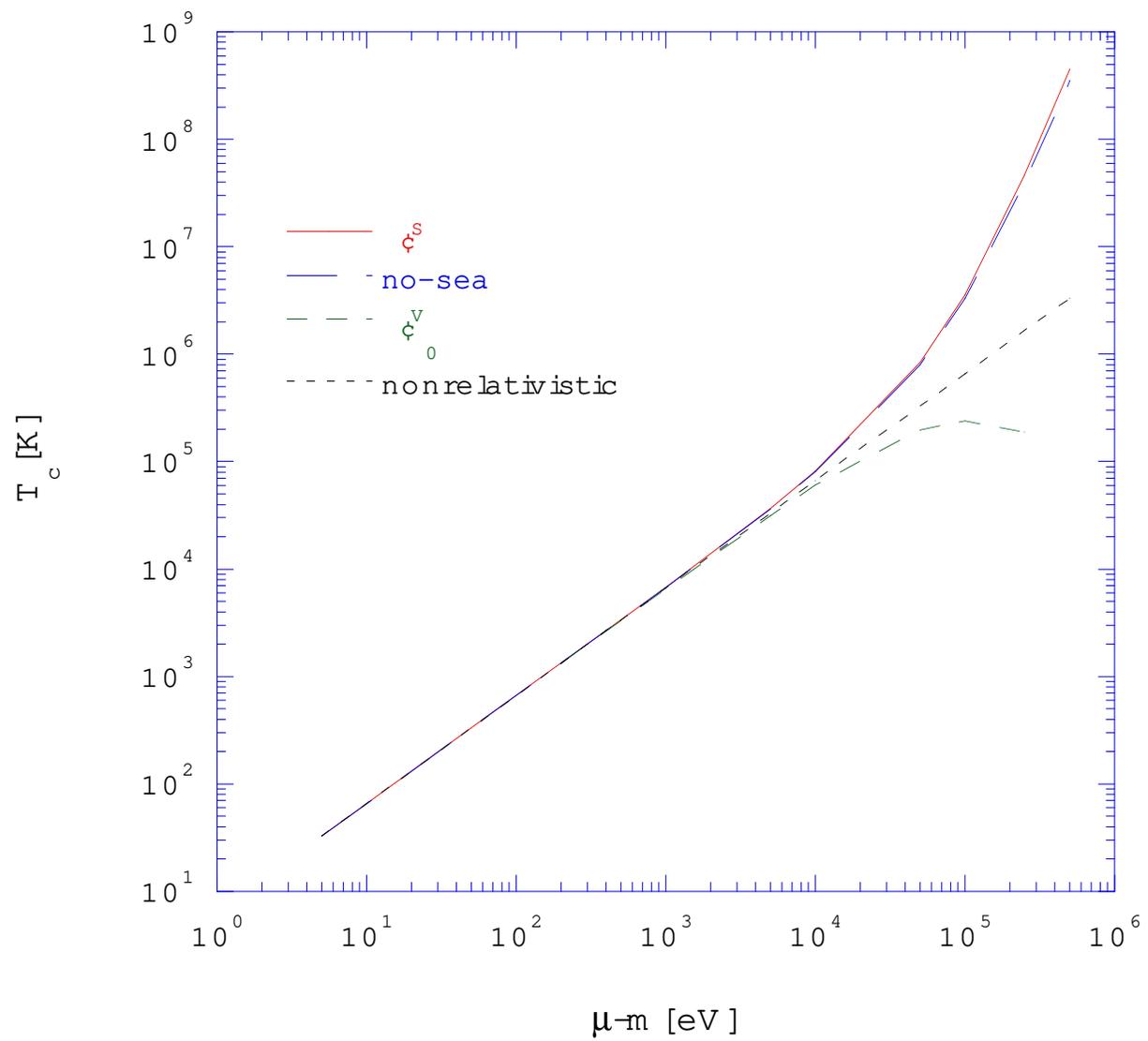
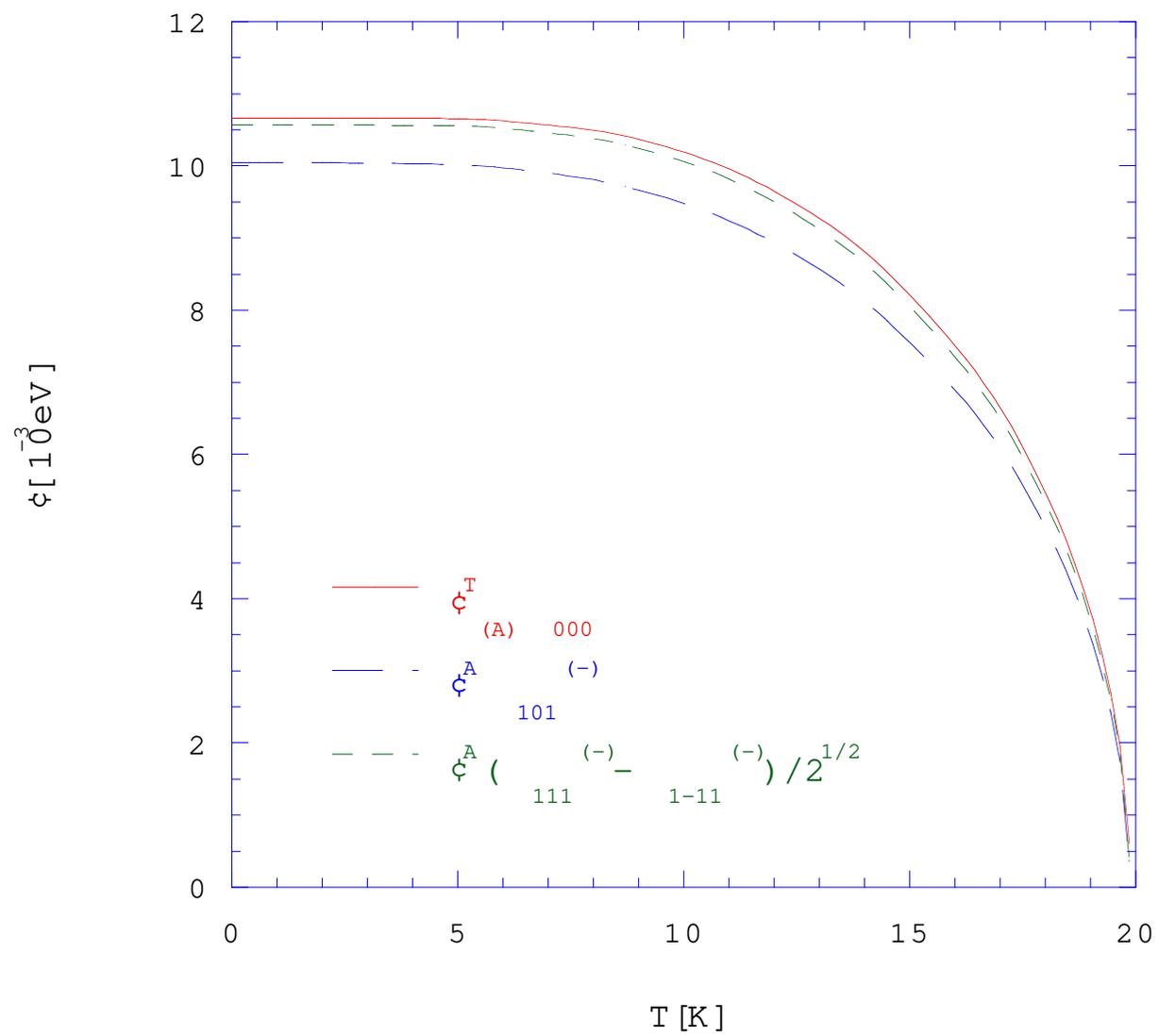
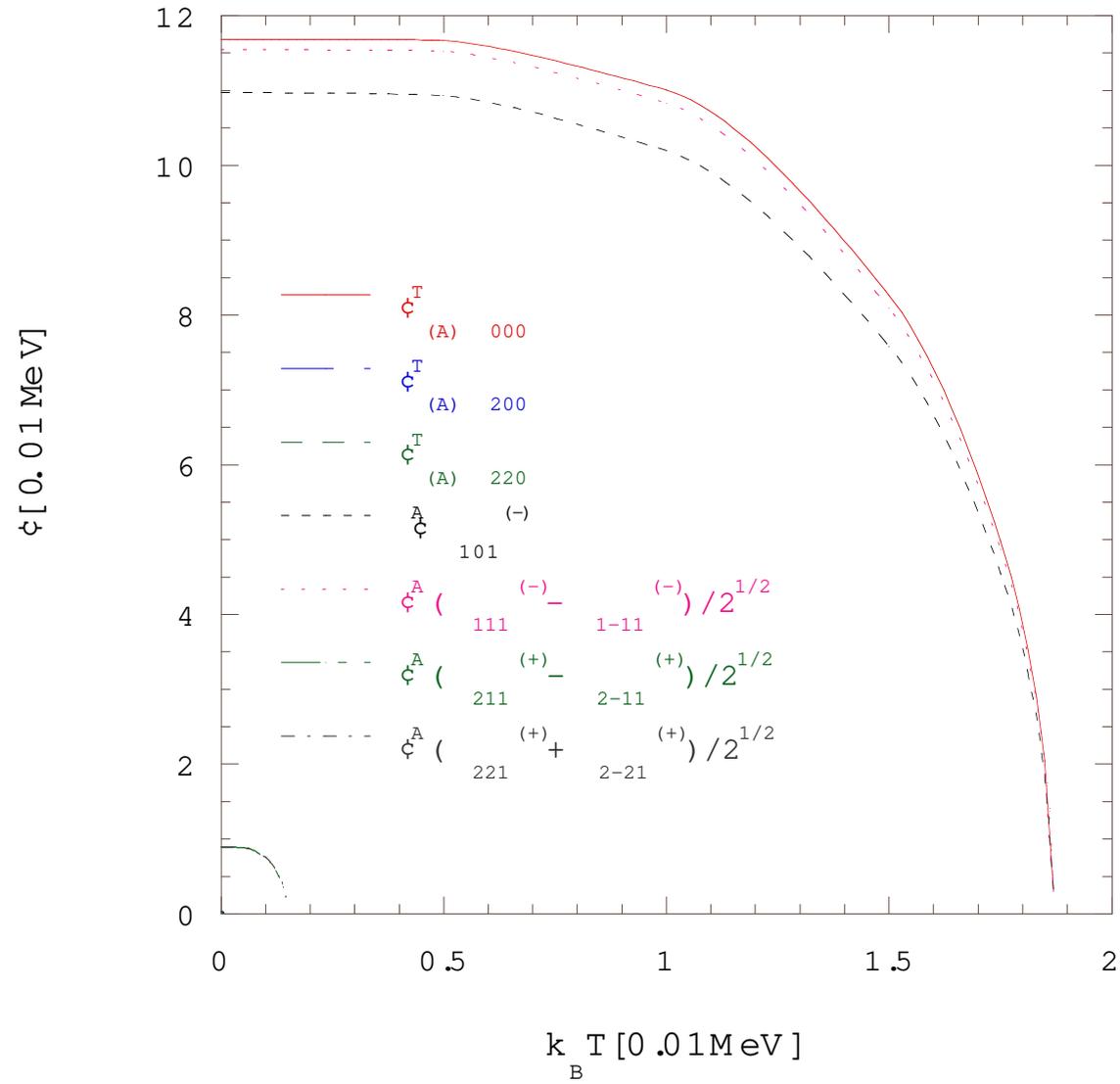
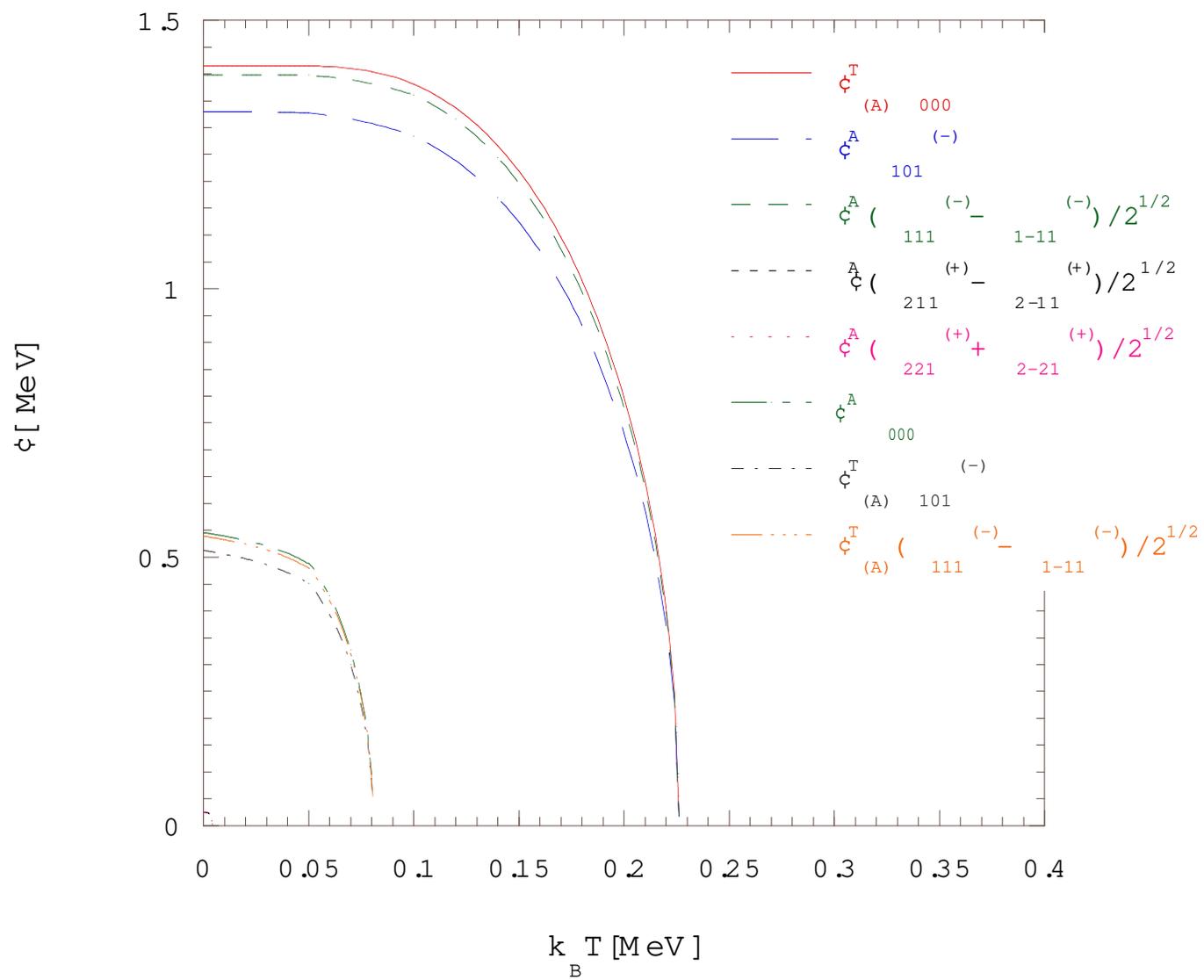
\end{document}